\documentclass[sigconf,10pt]{acmart}
\renewcommand\footnotetextcopyrightpermission[1]{} 
\renewcommand{\acmConference}{}
\renewcommand{\shortauthors}{}

\usepackage{tikz}
\usepackage{amsmath}
\usepackage{filecontents}
\usepackage{amsbsy}
\usepackage{algorithmic}
\usepackage{balance}
\usepackage{caption}
\usepackage{gensymb}
\usepackage{graphicx}

\usepackage{subfigure}
\usepackage{dblfloatfix} 
\usepackage{textcomp}
\usepackage{url}
\usepackage{xcolor}

\usepackage[english]{babel}
\usepackage{blindtext}

\usepackage{tikz}
\usepackage{amsmath,amssymb,amsfonts,mathrsfs}
\usepackage{latexsym}
\usepackage{algorithmic}
\usepackage{graphicx}
\usepackage{xcolor}

\usepackage{gensymb}
\usepackage{mathtools}
\usepackage{url}
\usepackage{balance}
\usepackage{xspace}
\usepackage{multirow}
\usepackage{tabularx}
\usepackage{makecell}
\usepackage{textcomp}
\usepackage{enumitem}
\usepackage{booktabs}
\usepackage{pifont}
\usepackage[vlined,ruled,linesnumbered]{algorithm2e}
\usepackage{pdfpages}
\usepackage{setspace}
\usepackage{svg}
\usepackage{siunitx} 
\usepackage{appendix}
\long\def\comment#1{}
\long\def\delete#1{}

\newcommand{\sysname}{\textsc{GaMi}\xspace}

\newcommand{\sssec}[1]{\vspace*{0.05in}\noindent\textbf{#1} }

\begin{document}

\title{\sysname: Geometry-Agnostic Material Identification via Cross-Modal Subtractive Disentanglement}

\author{Zhiwei Chen}
\authornote{Both authors contributed equally to this research.}
\orcid{0009-0004-8081-1360}
\affiliation{%
  \institution{UESTC}
  \city{Chengdu}
  \country{China}
}
\email{zhiwei.chen@std.uestc.edu.cn}

\author{Yijie Li}
\authornotemark[1]
\affiliation{%
  \institution{National University of Singapore}
  \country{Singapore}
}
\email{yijieli@nus.edu.sg}

\author{Yimo Zhang}
\affiliation{%
  \institution{UESTC}
  \city{Chengdu}
  \country{China}
}
\email{zym200336@163.com}

\author{Shiyun Shao}
\affiliation{%
  \institution{UESTC}
  \city{Chengdu}
  \country{China}
}
\email{shaosycd@gmail.com}

\author{Yichao Chen}
\affiliation{%
  \institution{Shanghai Jiao Tong University}
  \city{Shanghai}
  \country{China}
}
\email{yichao@sjtu.edu.cn}

\author{Dian Ding}
\affiliation{%
  \institution{Shanghai Jiao Tong University}
  \city{Shanghai}
  \country{China}
}
\email{dingdian94@sjtu.edu.cn}

\author{Liang Wang}
\affiliation{%
  \institution{Northwestern Polytechnical University}
  \city{Xi'an}
  \country{China}
}
\email{liangwang@nwpu.edu.cn}

\author{Haiwei Wu}
\affiliation{%
  \institution{UESTC}
  \city{Chengdu}
  \country{China}
}
\email{haiweiwu@uestc.edu.cn}

\author{Liwei Guo}
\affiliation{%
  \institution{UESTC}
  \city{Chengdu}
  \country{China}
}
\email{lwg@uestc.edu.cn}

\author{Jie Yang}
\affiliation{%
  \institution{UESTC}
  \city{Chengdu}
  \country{China}
}
\email{jie.yang@uestc.edu.cn}

\author{Xiaosong Zhang}
\affiliation{%
  \institution{UESTC}
  \city{Chengdu}
  \country{China}
}
\email{johnsonzxs@uestc.edu.cn}

\author{Yongzhao Zhang}
\authornote{Corresponding author.}
\affiliation{%
  \institution{UESTC}
  \city{Chengdu}
  \country{China}
}
\email{zhangyongzhao@uestc.edu.cn}

\begin{abstract}

Non-contact material identification enables adaptive interaction for embodied intelligence yet faces challenges from geometry-induced variations (e.g., orientation, shape, distance) and single-modality ambiguities. In this paper, we present \sysname,  a multimodal material identification system integrating mmWave and acoustic sensing to robustly operate under unconstrained geometric conditions. By leveraging the insight of shared geometric consistency between co-located bimodal sensors, \sysname employs an intra-sample cross-modal subtractive disentanglement framework. By semantically aligning modalities and subtracting the shared geometric context, it isolates intrinsic material features. 
Furthermore, \sysname incorporates inter-sample contrastive learning to correct the residual interference caused by cross-modal misalignment. Additionally, a pairing-based adaptation strategy between two modalities enables few-shot generalization across devices. Extensive evaluations on 20 materials show that \sysname achieves 95.2\% accuracy, outperforming single-modality baselines across unseen geometric conditions.

\end{abstract}

\maketitle
\thispagestyle{plain}
\pagestyle{plain}
\section{Introduction}
Wireless material identification aims to infer intrinsic object properties without physical contact, enabling applications such as liquid/food quality monitoring~\cite{viliquid,rfeats} and soil contamination detection~\cite{strobe}. 
Beyond these relatively static scenarios, embodied intelligence increasingly requires 
\textit{material awareness under various viewpoint changes}: robots should act reliably under continuously varying distance, orientation, and object geometry, while their decisions hinge on material properties. Such awareness directly supports adaptive interaction strategies, e.g., adjusting grasp force for fragile items~\cite{crackingegg}, planning manipulation trajectories for deformable materials~\cite{featuregeasping}, and avoiding objects that pose safety risks~\cite{riskavoiding}.

\begin{figure}[t]
	\centering
	\begin{minipage}[t]{0.64\linewidth}
		\centering
		\subfigure[Single-modal sensing fails in material identification with different geometries.]{%
			\includegraphics[width=\linewidth]{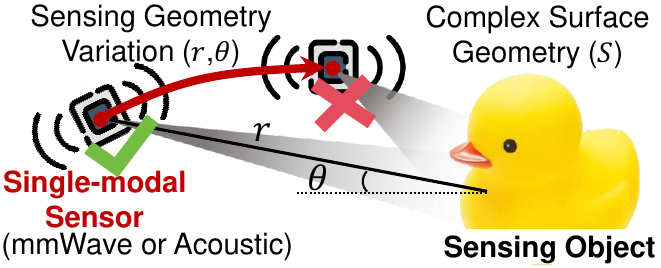}%
			\label{fig:intro1}%
		}\\[-3pt] 
		\subfigure[Bimodal sensing offers geometry-agnostic material identification.]{%
			\includegraphics[width=\linewidth]{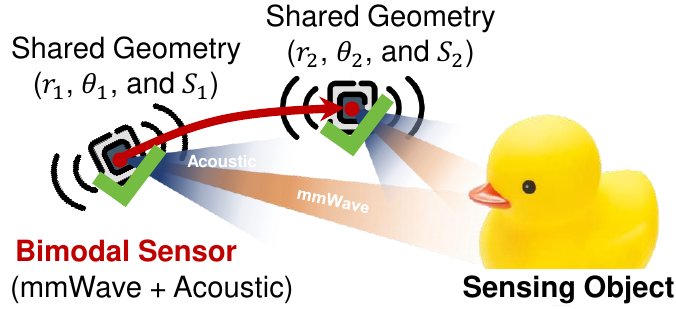}%
			\label{fig:intro2}%
		}
	\end{minipage}\hfill
	\begin{minipage}[t]{0.34\linewidth}
		\centering
		\subfigure[Cross-modal subtractive Disentanglement.]{%
			\includegraphics[width=\linewidth]{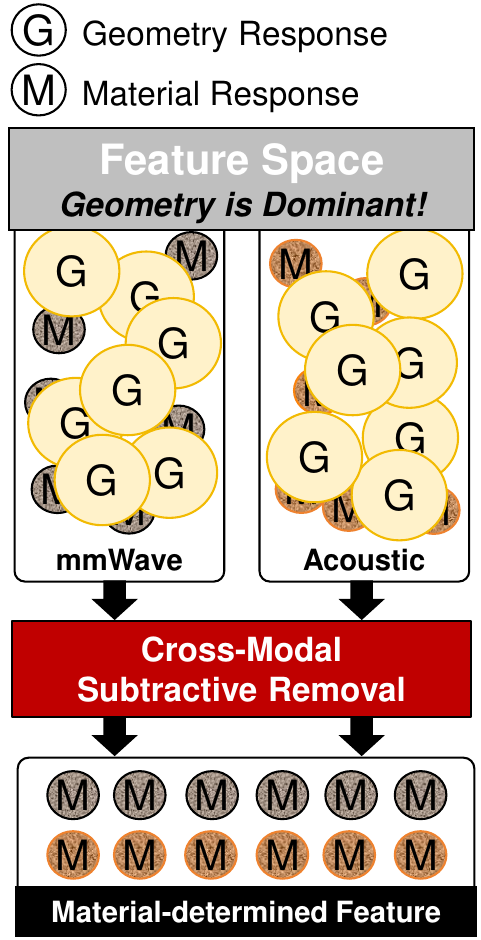}%
			\label{fig:intro3}%
		}
	\end{minipage}
	\vspace{-10pt} 
	\caption{Illustration of \sysname. It leverages cross-modal subtractive disentanglement framework to achieve geometry-agnostic material identification.}
	\label{fig:intro}
	\vspace{-15pt}
\end{figure}

Vision-based approaches~\cite{cv,cv1,cv2,cv3,cv4,cv5,cv6} are widely adopted due to camera availability, but they mainly capture surface appearance and can fail under camouflage, texture ambiguity, or illumination changes. Vibration-based approaches~\cite{viliquid,vibmilk,asliquid,hearliquid,akteliquid} analyze resonance excited by an external actuator, which typically requires physical contact or specialized containers. In contrast, RF-based solutions~\cite{rfray,intuwition,wimi,siwa,liquid,fgliquid,www,msense,mid,rfvibe} can probe intrinsic physical properties without contact. However, their received signals are often susceptible to channel-state variations induced by geometry (e.g., shape, orientation, distance), shown as Fig.~\ref{fig:intro1}. Prior work mitigates such geometry-induced interference by imposing constrained placement~\cite{rfvibe,wimi} or assuming standardized geometries~\cite{msense,mid}, which limits applicability in embodied settings where robots encounter diverse objects from varying viewpoints. These constraints motivate a central question: \textit{Can we enable non-contact material identification that remains reliable under substantial variations in complex surface geometry and sensing geometry?}

In this paper, we propose \sysname, a \textbf{bimodal} material identification system that integrates mmWave and acoustic sensing to achieve geometry-agnostic material recognition. Specifically, \sysname remains robust to variations in sensing geometry (distance $r$, orientation $\theta$) and surface geometry (shape $S$) (Fig.~\ref{fig:intro2}). Moreover, \sysname captures both \textit{electromagnetic} and \textit{mechanical} properties to resolve single modality's ambiguities and identify a broader range of materials. To achieve geometry-agnostic identification, \sysname is grounded in the insight of \textit{shared geometry consistency} of co-located sensors: despite distinct physical interactions, both modalities are governed by the identical geometry context of the object—the orientation $\theta$, distance $r$, and surface shape $S$ (shown as Fig.~\ref{fig:intro2}). Leveraging this, \sysname employs a novel \textbf{subtractive} design (Fig.~\ref{fig:intro3}) that filters out these shared context, thereby disentangling intrinsic material-centric representations regardless of geometry variations. In practice, realizing \sysname overcomes the following \textbf{challenges}:

\textit{\textbf{\#1.} \underline{Subtractive Disentanglement:} How to isolate material-centric features from dominant geometric features?}

The core challenge lies in the \textit{dominance of geometric features}, where variations in orientation, distance or shape are deeply entangled with and often overshadow the subtle material signatures. Traditional feature-joint multi-modal frameworks~\cite{radarcam,rcbevdet} often fail by reinforcing these geometric biases. Instead, our objective is fundamentally \textit{subtractive}: \sysname aims to explicitly identify and discard the shared geometry consistency to isolate intrinsic material representations. 

To realize this, \sysname introduces an \textbf{Intra-sample Subtractive Disentanglement Framework} to obtain an initial material-centric decomposition (Sec.~\ref{sec:method-disentanglement}). Specifically, the process begins with semantic alignment, followed by a cross-modal attention mechanism that calibrates the magnitude differences between modalities to ensure the validity of mathematical subtraction. By subtracting the aligned features corresponding to the shared geometry context, \sysname distills material-centric representations. We further refine this separation by enforcing an orthogonality and a reconstruction constraint, compelling the material and geometric representations to remain statistically independent. This design aligns closely with physical principles, where geometry is shared but material sensitivity differs, allowing the system to substantially reduce dominant geometric interference for each bimodal sample.

\textit{\textbf{\#2.} \underline{Residual Suppression:} How to suppress subtraction-induced residuals caused by cross-modal waveform misalignment?}

Although the intra-sample subtractive module provides an initial material-centric estimate, residual interference still remains due to cross-modal waveform misalignment.
This misalignment originates from the distinct sensing mechanisms: mmWave radar employs directional beamforming, whereas acoustic sensing is typically omnidirectional. 
Consequently, even under the same geometric scene, the geometry-related response patterns in target and environmental reflections are not perfectly matched across modalities, making clean cancellation difficult with a single sample alone.

To overcome this, \sysname introduces an \textbf{Inter-sample Contrastive Learning Misalignment Correction} (Sec.~\ref{sec:contra}). This strategy clusters features of the same material while separating those of different materials by leveraging material-centric features from multiple samples.
This explicitly penalizes sensitivity to waveform misalignment across modalities, compelling the model to learn consistent material features with smaller variance. Consequently, the non-aligned, location-dependent noise is filtered out, enhancing the purity and robustness of the final material representations. Notably, this process does not require precise metric coordinates in the dataset, thus reducing manual annotation effort.

\textit{\textbf{\#3.} \underline{Device Generalization:} How to ensure robust generalization across heterogeneous devices with limited labeled data?}

Different sensor implementations imprint unique hardware signatures on signals, varying in antenna/microphone configuration, sensitivity, and frequency response, etc. A model trained on a specific device may become overly reliant on hardware-specific features, causing performance degradation when deployed to a new device. Since recollecting large-scale datasets for every deployment is impractical, \sysname faces the challenge of cross-device generalization using only a few samples to minimize user effort.

To this end, \sysname adopts a \textbf{pairing-based cross-device adaptation} strategy (Sec.~\ref{sec:general}). Instead of recollecting large-scale new-device data, it expands the effective training diversity using limited target-device samples. Specifically, \sysname introduces \textit{sample-level pairing} to augment multimodal combinations from scarce target-device data, and \textit{feature-level pairing} to enrich geometry diversity under one-site calibration setting. Together, these two pairing mechanisms jointly improve both sample diversity and geometry diversity, enabling rapid adaptation for new devices.

We implemented \sysname using all off-the-shelf sensors, including an mmWave radar (TI IWR1843~\cite{TI_IWR1843}), with a pair of speaker and microphone.  Specifically, we conducted experiments on 20 common materials across multiple indoor environments. In overall-geometry split and strict unseen-geometry split evaluations (see Sec.~\ref{sec:evaluation_methodology}), \sysname achieves average identification accuracies of 95.2\% and 90.08\%, respectively, significantly outperforming the state-of-the-art baseline~\cite{mid}.
We also demonstrated cross-device generalization capability, \sysname maintains 91.01\% accuracy on new sensors with only one-site calibration for each material.

Our main contributions are summarized as follows:
\begin{itemize}[leftmargin=15pt, topsep=3pt]%

\item We propose \sysname, a robust multimodal material identification system that achieves geometry-agnostic recognition by integrating mmWave and acoustic bi-modality, while supporting the classification of a broader range of materials than single-modality approaches.

\item We design an intra-sample subtractive disentanglement network that isolates intrinsic material-centric features from dominant geometric artifacts.

\item We develop an inter-sample contrastive learning-based method to learn clean material features by suppressing residual interference caused by cross-modal waveform misalignment. 

\item We introduce a pairing-based cross-device adaptation scheme to enable few-shot generalization to new devices with minimal data collection.

\item We implement \sysname and conduct extensive evaluations on 20 common materials, demonstrating 95.2\% accuracy under varying distances, orientations, and shapes.
\end{itemize}

\section{Related Work}
\subsection{Material Identification}
\sssec{Single-modality approaches.} Most existing solutions are based on a single sensing modality.

\noindent\textit{1) Vision-based approaches.} Vision-based solutions utilize cameras to extract texture features~\cite{cv,cv1,cv2,cv3}. However, they are fundamentally limited to surface appearance and are susceptible to illumination variations and environmental conditions~\cite{cv4,cv5}. Moreover, cameras cannot perceive intrinsic properties, thus are easily deceived by surface camouflage or visual imitation~\cite{cv6}, leading to unreliable identification.

\noindent\textit{2) Vibration-based approaches.} Vibration-based methods infer mechanical properties (e.g., density and elasticity) by analyzing responses to external actuation or acoustic excitation using commodity devices like smartphones or earphones~\cite{viliquid,vibmilk,asliquid,hearliquid}.  Akte-Liquid~\cite{akteliquid} explores over-the-air sensing via amplitude-frequency responses but requires strict alignment. While effective for liquid estimation, these solutions typically impose strict \textbf{physical constraints}, such as requiring direct contact or specific containers.

\noindent\textit{3) RF-based approaches.} RF signals can probe intrinsic dielectric properties non-intrusively. \textit{Contact-based approaches}~\cite{radarcat,rfiq,tagtag,rfeats,twinleak} leverage coupling effect, offering precision but lack flexibility due to physical contact and prior instrumentation. Conversely, \textit{non-contact solutions} ~\cite{rfray,intuwition,wimi,siwa,liquid,fgliquid,www,mid} enable standoff sensing but are highly sensitive to \textbf{geometry-induced variations}. The object shape, orientation, and distance often dominate the received signal. Consequently, existing works often rely on large antenna arrays~\cite{msense} or assume standardized geometries~\cite{mid} to improve accuracy, limiting their robustness in unconstrained environments.

\sssec{Multi-modality approaches.} Existing single-modality approaches face a fundamental trade-off between deployment flexibility and sensing robustness, motivating the exploration of multi-modal solutions. RFVibe~\cite{rfvibe}, combines mmWave and acoustic sensing but relies on detecting acoustic-induced vibrations. This approach essentially measures a single physical response and requires high-power excitation with strict positioning constraints, introducing undesirable noise and limiting practical utility.
\subsection{Multi-Modal Learning}
Beyond material sensing, multi-modal learning~\cite{mmsurvey} has illustrates strong performance by leveraging \textbf{unified information} across domains such as vision--language learning~\cite{vit,clip,albef}, autonomous driving~\cite{rcbevdet,radarcam}, human activity recognition~\cite{ouyang2022cosmo}, and plant sensing~\cite{hydra,liu2025proteus,ding2024cost}. Most frameworks align modalities into a unified semantic space to capture common factors. However, directly applying alignment-based fusion to material identification is non-trivial: naive alignment or fusion can amplify dominant geometric and environmental biases instead of isolating intrinsic material cues. Existing disentanglement methods~\cite{decur,demo} often assume separable factors (e.g., sketch vs. appearance~\cite{decur}), whereas material signatures are subtle and strongly entangled with geometry.

\sssec{$\blacksquare$ Summary of Difference.} 
In our work, \sysname leverages mmWave and acoustic bimodal signals to capture both electromagnetic and mechanical responses. 
Rather than aligning or concatenating multimodal features, \sysname proposes a \textbf{subtractive disentanglement framework} that explicitly filters out geometry components shared across both modalities. By isolating intrinsic material signatures from dominant geometric factors such as shape, orientation, and distance, \sysname enables robust material identification under diverse object geometries and unconstrained sensing configurations.


\section{Background and Challenges}
\subsection{Physical Basis of Material Response}
\label{subsec:basis}
mmWave radar and acoustic sensing probe fundamentally different material properties. mmWave interactions are governed by dielectric permittivity, whereas acoustic are dictated by mechanical impedance.

\sssec{Electromagnetic Response.}
mmWave interactions with materials are governed by the complex dielectric
permittivity $\varepsilon_m$. In practice, the observed mmWave response results from a spatial integration over the object surface, jointly determined
by the local incident orientation, propagation distance, and surface geometry~\cite{mmrespone}.
Where the effective mmWave reflection response\footnote{Frequency dependence is omitted for simplification since relative material contrast and geometric effects dominate within our operational bandwidth.} $E_{mm}$ from the object is:
\begin{equation}
E_{mm}(\boldsymbol{\theta}, \boldsymbol{r}, S, \varepsilon_m)
=
\int_{S}
\left|
\frac{\cos\boldsymbol{\theta} - \sqrt{\varepsilon_m - \sin^2\boldsymbol{\theta}}}
{\cos\boldsymbol{\theta} + \sqrt{\varepsilon_m - \sin^2\boldsymbol{\theta}}}
\right|^2
\, \frac{1}{{\boldsymbol{r}}^2}
\, dS
\label{eq:mmwave_reflection}
\end{equation}
where $S$ denotes the object shape, and $\boldsymbol{r}$ and $\boldsymbol{\theta}$ represent the sets of point-wise sensor-to-surface \textit{distances} and \textit{orientation} (defined between the incident wave direction and the surface normal), respectively, as illustrated in Fig.~\ref{fig:intro}.

\sssec{Mechanical Response.} Relatively, acoustic sensing captures the mechanical properties of materials via their acoustic impedance $Z_m = \rho_m c_m$, which is determined by density $\rho_m$ and the propagation sound speed $c_m$~\cite{acrespone}. The effective acoustic reflection response $E_{ac}$ is also spatially modulated by the object's shape $\mathbf{S}$, orientation $\theta$ and distance $r$:
\begin{equation}
E_{ac}(\boldsymbol{\theta}, \boldsymbol{r}, S, Z_m')
=\int_{S}
\left|\frac{Z_m' \cos\boldsymbol{\theta} - \sqrt{1 - k^2 \sin^2\boldsymbol{\theta}}}{Z_m' \cos\boldsymbol{\theta} + \sqrt{1 - k^2 \sin^2\boldsymbol{\theta}}}\right|^2 \frac{1}{{\boldsymbol{r}}^2}
\,dS
\label{eq:acoustic_reflection}
\end{equation}
where we define the normalized material impedance $Z_m'= Z_m / Z_\mathrm{air}$ and $k = c_\mathrm{air}/c_m$ is the ratio of sound speeds.

\begin{figure}[H]  
	\centering
	\vspace{-10pt} 
	\subfigure[{\small Property distributions across different materials.}]{%
		\includegraphics[width=0.46\linewidth]{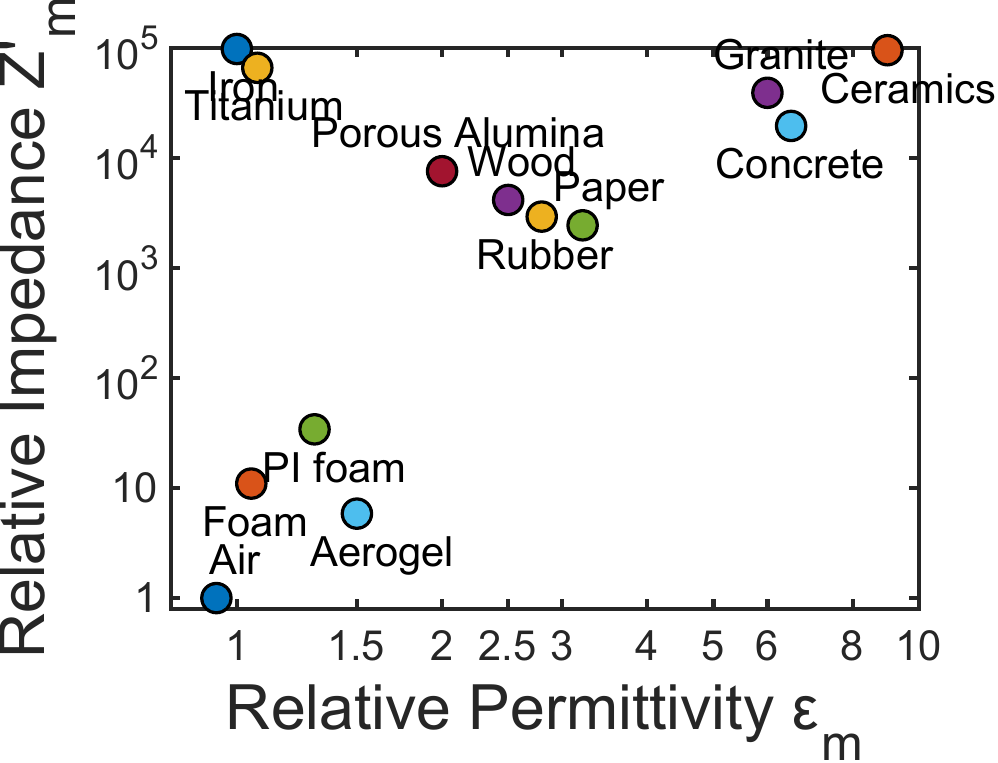}%
		\label{fig:sub2}%
	}\hfill
	\subfigure[{\small Ambiguity with single modality view.}]{%
		\includegraphics[width=0.51\linewidth]{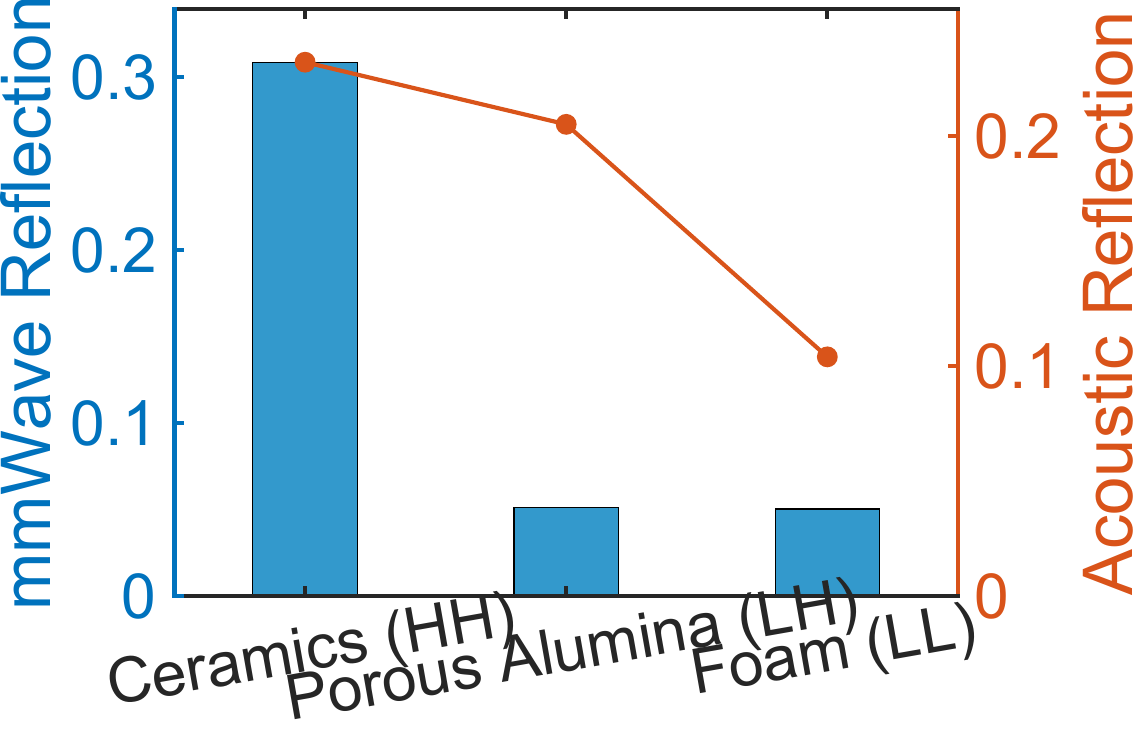}%
		\label{fig:sub1}%
	}
	\vspace{-15pt} 
	\caption{Motivation for multimodal sensing: resolving single-modality ambiguity by complementing electromagnetic and mechanical responses.}
	\label{fig:combined}
	\vspace{-15pt} 
\end{figure}

\sssec{$\blacksquare$ Motivation: why does single modality sensing fall short?} Prior works~\cite{msense,mid} demonstrated material identification using mmWave sensing alone, but they are inherently limited when materials exhibit similar dielectric properties. As shown in Fig.~\ref{fig:sub2} (data sourced from the COMSOL~\cite{comsol}), many materials overlap in a single-property view. To validate this ambiguity, we ran COMSOL simulations with representative material parameters (e.g., ceramics, porous alumina, and foam). As shown in Fig.~\ref{fig:sub1}, mmWave sensing struggles to distinguish foams and porous alumina because both have low dielectric permittivity ($\varepsilon_m$) and yield similarly weak reflections. Conversely, acoustic sensing can be ambiguous for materials with similarly high normalized acoustic impedance ($Z_m'$), such as ceramics and porous alumina. 
These results motivate fusing electromagnetic and mechanical perspectives to resolve single-modality ambiguity and differentiate a broader range of materials.

\subsection{Geometry-Entangled Sensing}
\label{sec:shared-geo}
Although multi-modality provides complementary physical properties, the received signals are not pure material signatures. Instead, they are heavily modulated by the object's state (i.e., geometry), sensing configuration (e.g., antenna/microphone response), and environmental interference. We model the real received mmWave ($R_{mm}$) and acoustic signals ($R_{ac}$) regarding specific devices as:

\begin{align}
R_{mm}
&=
\Big(
E_{mm}(\boldsymbol{\theta}, \boldsymbol{r}, S, \varepsilon_m)
+
\sum \alpha^{mm} \, e^{-j \psi^{mm}}
\Big)
\cdot G_{\mathrm{sys}}^{mm}
\label{eq:received_signal} \\
R_{ac}
&=
\Big(
\underbrace{E_{ac}(\boldsymbol{\theta}, \boldsymbol{r}, S, Z_m')}_{\text{Target Reflection}}
+
\underbrace{\sum \alpha^{ac} \, e^{-j \psi^{ac}}}_{\text{Multipath Reflection}}
\Big)
\cdot 
\underbrace{G_{\mathrm{sys}}^{ac}}_{\text{Device}}
\label{eq:acoustic_received}
\end{align}
As modeling in Sec.~\ref{subsec:basis}, the reflection response coefficient \(E_{mm}\) and \(E_{ac}\) are determined jointly by material properties and geometric context.
The summation term represents contributions from surrounding reflection captured by each sensor, where \(\alpha\) denotes the attenuation coefficient. Finally, \(G_{\mathrm{sys}}\) accounts for device-specific factors, including transmit power, the number of antennas or transducers, gains, and frequency responses. 

\sssec{$\blacksquare$ Key insight: Shared Geometric Consistency.} Despite the distinct physical mechanisms (electromagnetic vs. mechanical), comparing Eq.~\ref{eq:received_signal} and Eq.~\ref{eq:acoustic_received} reveals a critical insight: 
When acoustic and mmWave sensors are \textbf{co-located}, both modalities capture nearly the same geometric context (i.e., orientation $\boldsymbol{\theta}$, distance $\boldsymbol{r}$, shape $S$) from identical object. This motivates \sysname to disentangle intrinsic material features (i.e., $\varepsilon_m, Z_m'$ ) by explicitly \textbf{subtracting these shared geometric effects}, thereby achieving geometric-agnostic material identification under dynamic conditions.
\begin{figure}[t]
	\centering
    \vspace{-10pt}
	\subfigure[mmWave reflection under impact of orientation $\theta$, sensing distance $r$, and surface shapes $\mathcal{S}$.%
	\label{fig:mmwave_group}]{%
		\centering
		\includegraphics[width=0.33\linewidth]{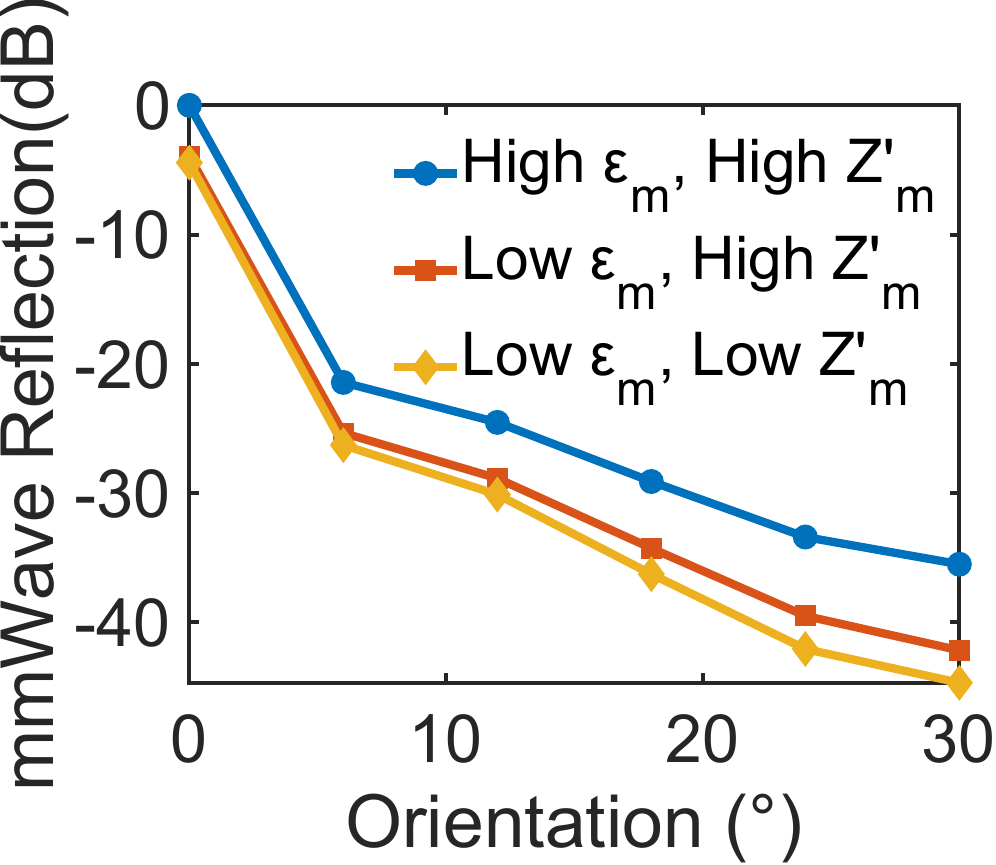}\hfill
		\includegraphics[width=0.33\linewidth]{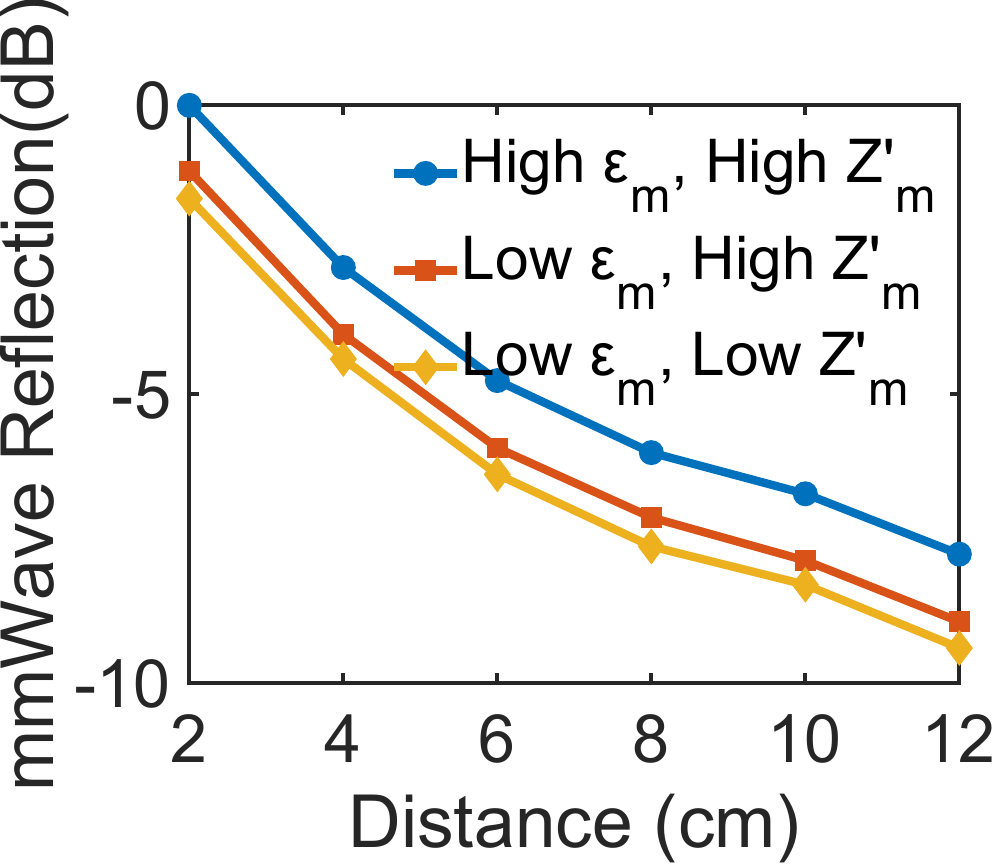}\hfill
		\includegraphics[width=0.32\linewidth]{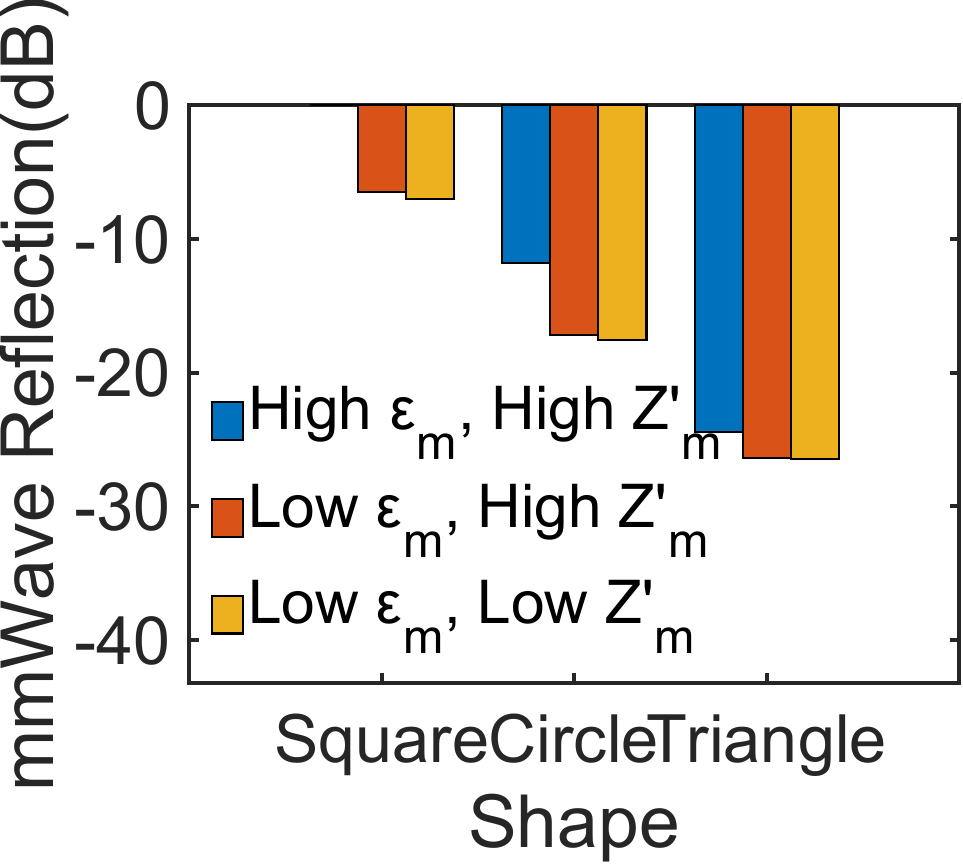}%
	}
	\vspace{-10pt}
	
	\subfigure[{Acoustic reflection under impact of orientation $\theta$, sensing distance $r$, and surface shapes $\mathcal{S}$.}%
	\label{fig:acoustic_group}]{%
		\centering
		\includegraphics[width=0.32\linewidth]{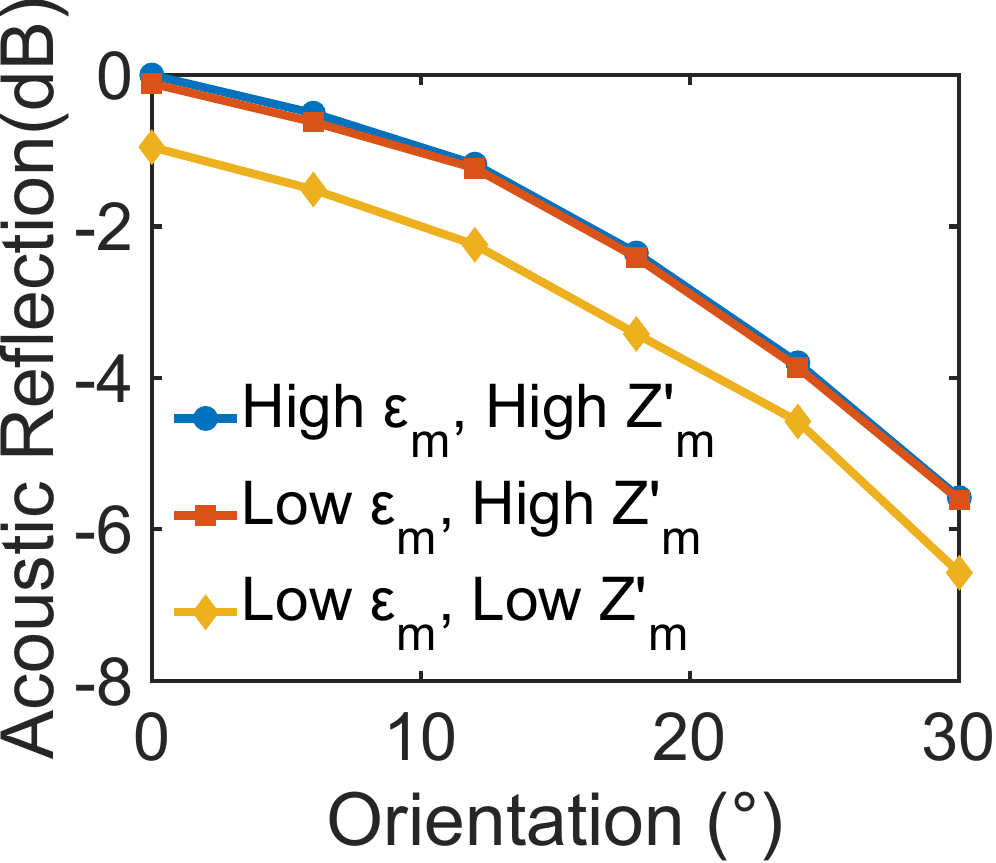}\hfill
		\includegraphics[width=0.33\linewidth]{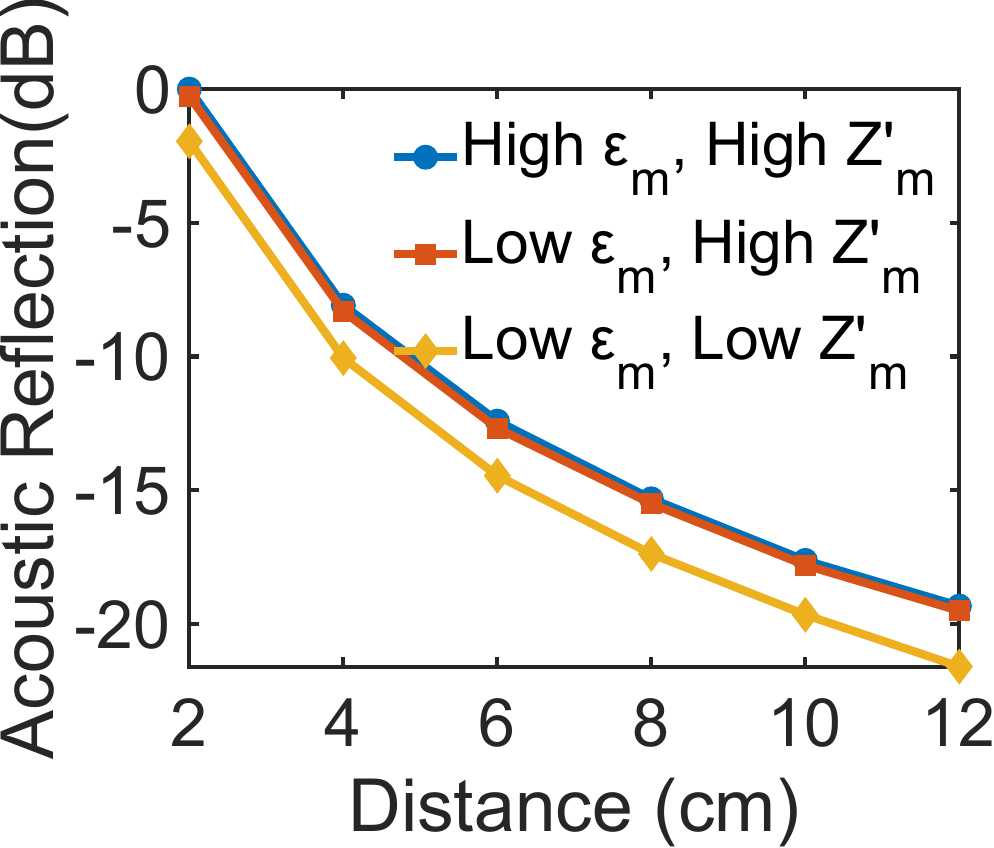}\hfill
		\includegraphics[width=0.31\linewidth]{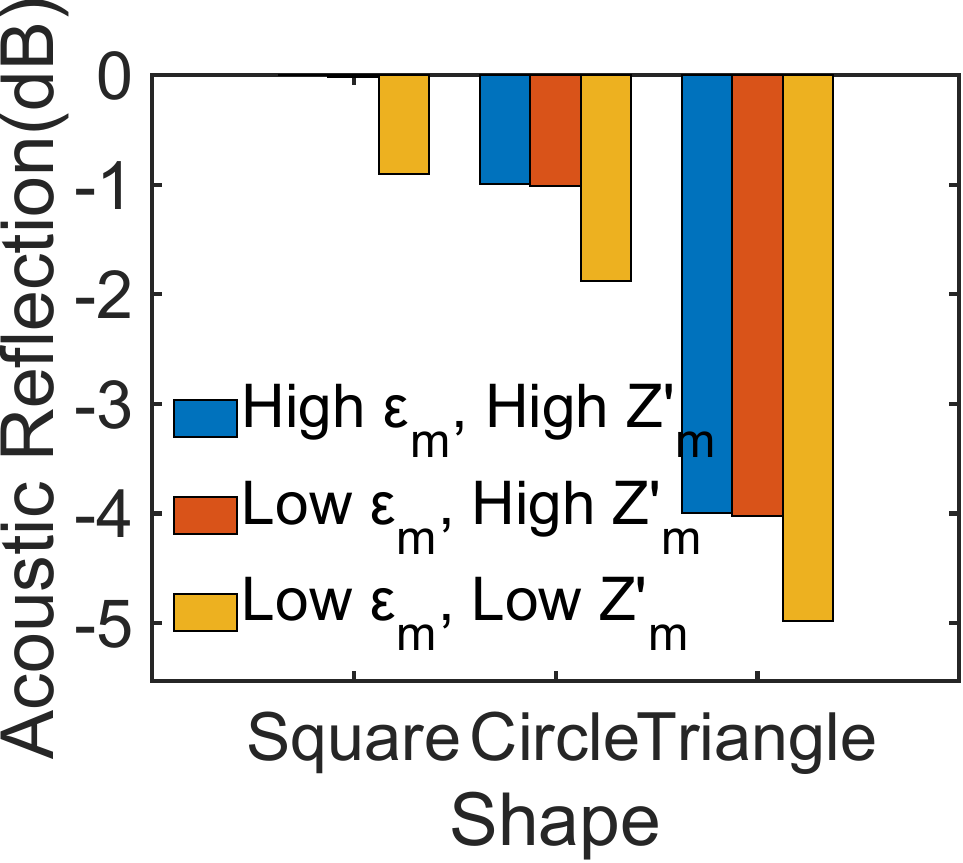}%
	}
	\vspace{-15pt}
	\caption{Comparison of mmWave and acoustic reflections under different geometric contexts.}
	\label{fig:all_multimodal}
	\vspace{-20pt}
\end{figure}

\subsection{Challenges}
\label{sec:simulationchallenges}
Realizing  subtractive disentanglement and robust material identification is non-trivial due to the following challenges:
\begin{figure*}[t]
	\centering
	\includegraphics[width=\linewidth]{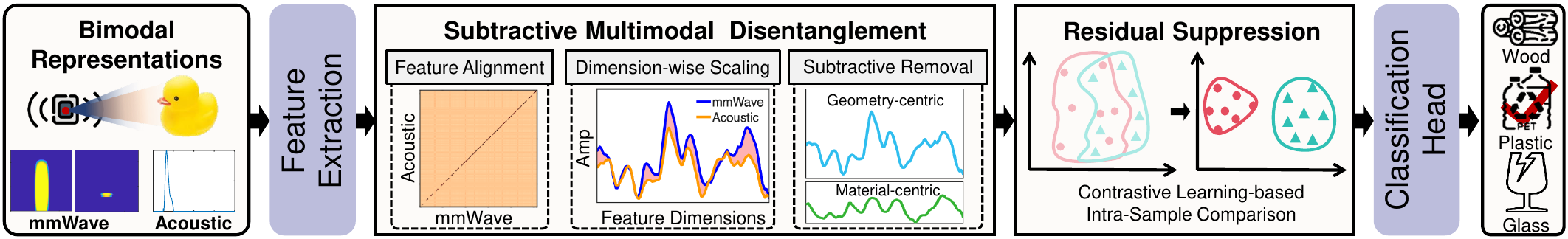}
	\vspace{-20pt}
	\caption{Overview of \sysname, which comprises four main components.
		Bimodal representations suppresses multipath interference;
		Subtractive multimodal disentanglement removes shared geometric factors to obtain material-centric representations; A contrastive learning-based framework is used to correct residual cross-modal waveform misalignment; and a head for final material classification.}
	\label{fig:overview}
	\vspace{-10pt}
\end{figure*}
\sssec{1) Geometry dominance and feature entanglement.} Intrinsic material signatures are deeply entangled with geometric information. Our simulations in Fig.~\ref{fig:all_multimodal} reveal that signal variance induced by geometric variations (i.e., orientation $\boldsymbol{\theta}$, distance $\boldsymbol{r}$, and surface shape $S$) often dwarfs the variance caused by material. This dominance can easily mask the subtle material signatures, creating a barrier to robust feature isolation. \sysname addresses this challenge in Sec.~\ref{sec:method-disentanglement}.

\sssec{2) Cross-modal waveform misalignment.} Although both modalities observe the same target from the same viewpoint, their propagation waveforms are not perfectly aligned due to distinct sensing mechanisms (e.g., the misalignment between mmWave beamforming $G_{\mathrm{sys}}^{mm}$ and quasi-spherical acoustic propagation $G_{\mathrm{sys}}^{ac}$). This misalignment affects not only in environmental multipath components, but also in the target reflections themselves.
As a result, the cross-modal subtractive module may retain subtraction-induced residuals, which is comparable to or even stronger than the material signatures.
To address this issue, \sysname introduces a contrastive-learning-based correction module in Sec.~\ref{sec:contra} to mitigate residual cross-modal misalignment.

\sssec{3) Device heterogeneity.} The device factors $G_{\mathrm{sys}}$, e.g., including the heterogeneity between two mmWave sensors ($G^{mm}_{\mathrm{sys,1}}$ and $G^{mm}_{\mathrm{sys,2}}$) and two acoustic sensors ($G^{ac}_{\mathrm{sys,1}}$ and $G^{ac}_{\mathrm{sys,2}}$), vary across hardware due to configurations in transmit power, antenna count, sensitivity, and frequency response. These hardware dependent factors induce distribution shifts in the learned features, which can significantly degrade generalization to new devices. This motivates \sysname to adopt paired-data cross-device adaptation (Sec.~\ref{sec:general}) to align the feature space with limited new samples. 

\section{\sysname Design}
In this section, we introduce \sysname, a multi-stage subtractive multimodal framework (as shown in Fig.~\ref{fig:overview}) that learns material-centric representations using mmWave and acoustic modalities. \sysname first constructs geometry-aware representations (Sec.~\ref{sec:method-geo}), then performs subtractive multimodal disentanglement to filter out the shared geometry information (Sec.~\ref{sec:method-disentanglement}). Next, it applies contrastive learning to correct residual cross-modal waveform misalignment (Sec.~\ref{sec:contra}) and finally performs supervised material classification (Sec.~\ref{sec:cls}). 
\subsection{Bimodal Representations}
\label{sec:method-geo}
Both mmWave and acoustic sensing capture geometry-aware representations from an identical target object.

\begin{figure}[t]
	\subfigure[{\small mmWave sensing}]{%
		\centering
		\includegraphics[width=0.68\linewidth]{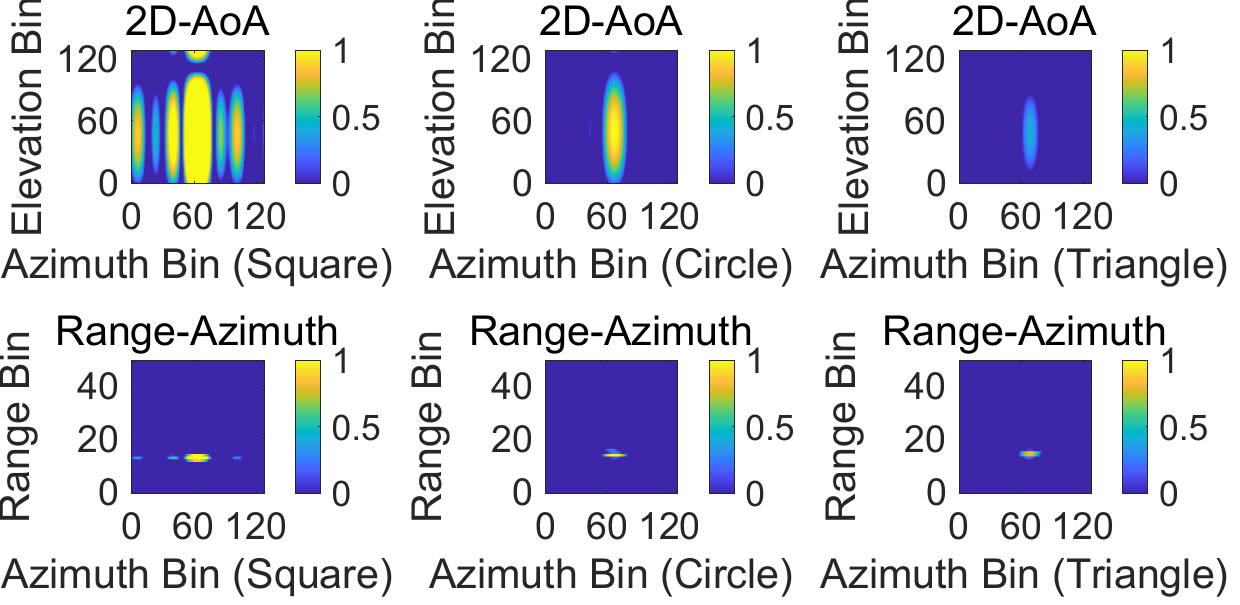}%
		\label{fig:mm}%
	}\hfill
	\subfigure[{\small Acoustic sensing}]{%
		\centering
		\includegraphics[width=0.3\linewidth]{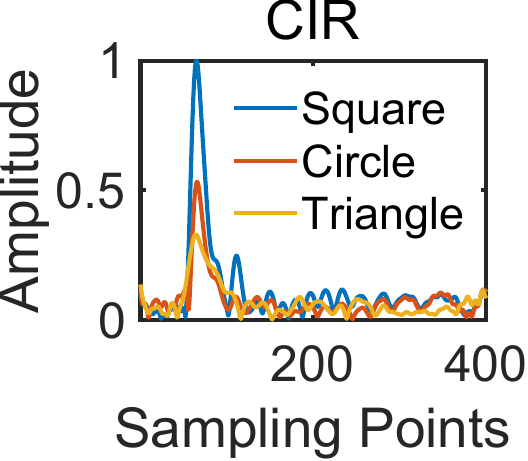}%
		\label{fig:ac}%
	}
	\vspace{-15pt} 
	\caption{Capturing geometry context using mmWave representations (2D-AoA and Range-azimuth heatmap) and acoustic representations (CIR).}
	\label{fig:CapturingGeometryInformation}
	\vspace{-15pt} 
\end{figure}

\sssec{MmWave Sensing.} We leverage FMCW signals to capture geometric details by estimating target distance via frequency differences and Angle-of-Arrival (AoA) via phase shifts across antennas. To fully characterize the geometry, we generate two representations:

\begin{itemize}[leftmargin=15pt]
\item \textit{1) 2D-AoA Heatmap:} We utilize a dual-channel virtual antenna array to capture spatial information. By processing phase vectors across these channels, we estimate the 2D angular spectrum at the target range. This heatmap encodes direction, azimuth and elevation.

\item \textit{2) Range-Azimuth Heatmap:} To capture spatial layout and shape, we compute the energy distribution on the horizontal plane. By stacking azimuth spectra computed at each range bin, the heatmap visualizes spatial energy dispersion, which reflects the target's shape and position. 
\end{itemize}
Together, these representations provide a comprehensive geometric profile. As shown in Fig.~\ref{fig:mm}, distinct shapes (square, circle, and triangle) exhibit unique reflection signatures: square targets produce wider angular spreads with multiple lobes, whereas circular and triangular targets yield more compact, concentrated reflections.

\sssec{Acoustic Sensing.} For acoustic sensing, we estimate the acoustic Channel Impulse Response (CIR) by transmitting broadband Zadoff–Chu (ZC) sequences~\cite{zcgood2,acsurvey}. Leveraging the ZC sequence's ideal autocorrelation properties (i.e., zero sidelobes), we achieve high-resolution temporal profiling. The propagation delay of the primary peak characterizes the target distance, while the overall amplitude and profiling shape contains geometric details, as illustrated in Fig.~\ref{fig:ac}, where different object shapes exhibit distinct CIR responses.

\begin{figure}[t]
    \centering
    \begin{minipage}[b]{0.49\linewidth}
        \centering
        \includegraphics[width=0.43\linewidth]{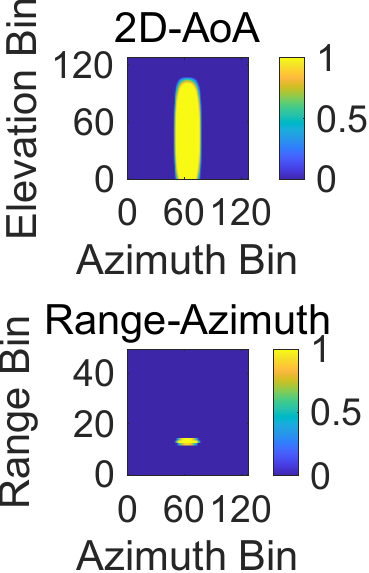}%
        \hfill
        \raisebox{14pt}{\includegraphics[width=0.50\linewidth]{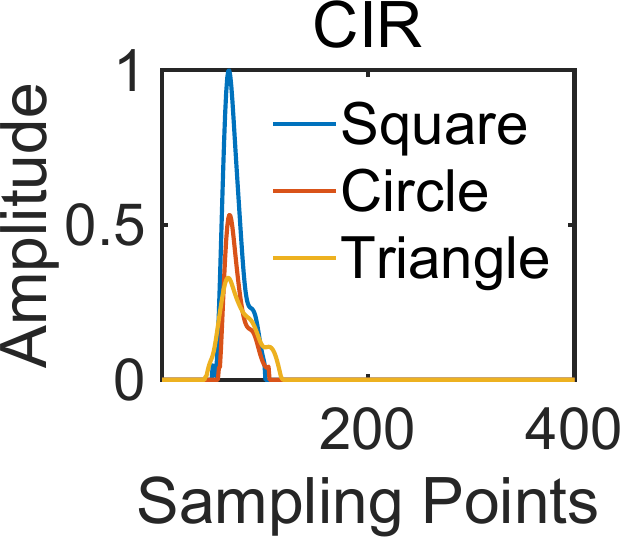}}%
        \vspace{-10pt}
        \caption{Suppressing multipath interference.}  
        \label{fig:suppress_multipath}
    \end{minipage}\hfill
    \begin{minipage}[b]{0.47\linewidth}
        \centering
        \includegraphics[width=\linewidth]{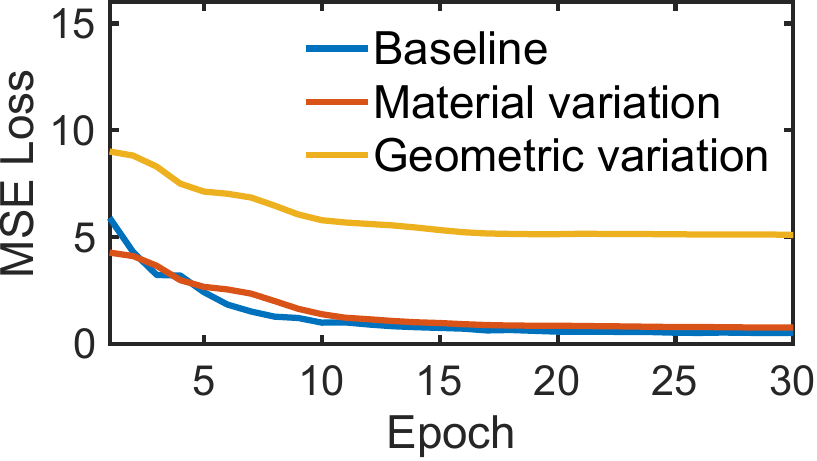}
        \vspace{-20pt}
        \caption{Cross-modal reconstruction error (MSE).}
        \label{fig:cross_modal_mse}
    \end{minipage}
    \vspace{-10pt}
\end{figure}

\sssec{Suppressing Multipath Interference.} 
In real-world settings, both modalities are susceptible to environmental multipath from the surroundings. To address this, we exploit the time-of-flight principle: since the direct reflection from the target\footnote{We assume the object being identified is closer to the receiver than the surrounding objects.} represents the shortest path and arrives earlier than multipath reflections.
Accordingly, for mmWave, we implement a peak-based filtering strategy to isolate the dominant target reflection while suppressing low-magnitude environmental clutter. This effectively reduces background interference while preserving the target-dominant structure in the mmWave inputs. 
Similarly, for acoustic sensing, we identify the earliest significant peak as the direct line-of-sight response and filter out subsequent multipath-induced tails.
Fig.~\ref{fig:suppress_multipath} show such pre-processing effectively concentrates signal energy on the target. We then feed the filtered signals, denoted as $x_{mm}$ and $x_{ac}$, into the network. 
Note that while this step reduces multipath interference, subtle modality-specific environmental interference and system noise still persist. Further suppression is addressed in Sec.~\ref{sec:contra}.

\subsection{Subtractive Multimodal Disentanglement}
\label{sec:method-disentanglement}
Unlike traditional multi-modal fusion that aims to aggregate all information, the goal of \sysname is fundamentally \textit{subtractive}: to explicitly filter out shared geometric information and isolate the intrinsic material features.

\subsubsection{Dominant Shared Geometry Verification.}\label{sec:DominantShared GeometryVerification}
Implementing the subtractive disentanglement is non-trivial due to the \textbf{dominance} of geometry-induced variations, as illustrated in Fig.~\ref{fig:all_multimodal}. Consequently, naive models tend to overfit these dominant geometric proxies, failing to capture the subtle, intrinsic material signatures. To empirically verify this entanglement, we designed a cross-modal reconstruction experiment where a lightweight network is trained to calculating Mean Square Error (MSE) in predicting acoustic CIRs solely from mmWave features under three controlled conditions:
(a) \textit{Baseline:} Identical geometry and materials;
(b) \textit{Material Variation:} Identical geometry but different materials;
(c) \textit{Geometric Variation:} Identical materials but different geometric configurations.

As illustrated in Fig.~\ref{fig:cross_modal_mse}, the MSE loss shows that the MSE for (b) converges to a close low level to the Baseline (a). This convergence implies that the cross-modal mapping remains effective even when materials change, suggesting the shared features are material-agnostic. In contrast, (c) exhibits a higher loss than baseline, which provides compelling evidence that geometric changes significantly interfere with material identification. This also illustrates that simple classification networks tend to fit dominant geometry and fail to disentangle marginal material-specific cues. The implementation of lightweight network is detailed in Appendix.~\ref{app:lightweightnetwork}

\begin{figure}[t]
	\centering
	\includegraphics[width=\linewidth]{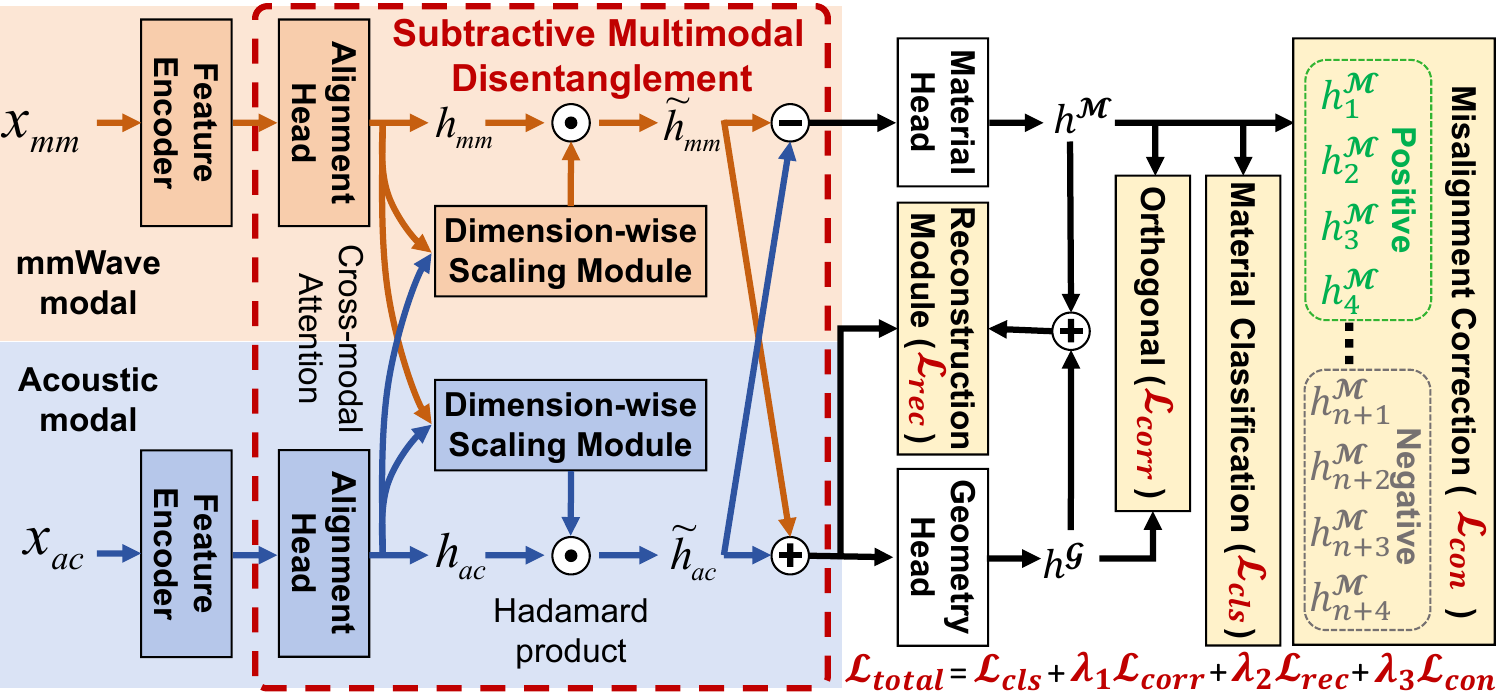}
	\vspace{-20pt}
	\caption{Material feature disentanglement framework.}
	\vspace{-15pt}
    \label{fig:disentanglement}
\end{figure}
\subsubsection{Disentanglement Framework}
\label{sec:disent}
In this section, we show an \textbf{intra-class subtractive disentanglement framework}, as shown in Fig.~\ref{fig:disentanglement}, to filter out geometric-induced noises. We achieve this through a multi-stage process:

\sssec{Feature Extraction.} 
We employ feature encoder $\Phi_F(\cdot)$, implemented with ResNet backbone~\cite{resnet,1dres}, to extract modality-specific features from each sensing modality. In the learned feature space, the representation can be approximately decomposed as $f = \Phi_F(x) \approx g + m + \epsilon$ for both acoustic and mmWave modalities, where $g$ denotes geometry information, $m$ represents material information, and $\epsilon$ captures measurement noise. 
Specifically, for mmWave sensing has two inputs, we concatenate the two extacted feature representations to form a unified mmWave feature vector.

\begin{figure*}[ht]  
	\centering
    \vspace{-10pt}
	\subfigure[{\small The correlation between $\mathbf{h}_{mm}$ and $\mathbf{h}_{ac}$.}]{
		\centering
		\includegraphics[width=0.18\linewidth]{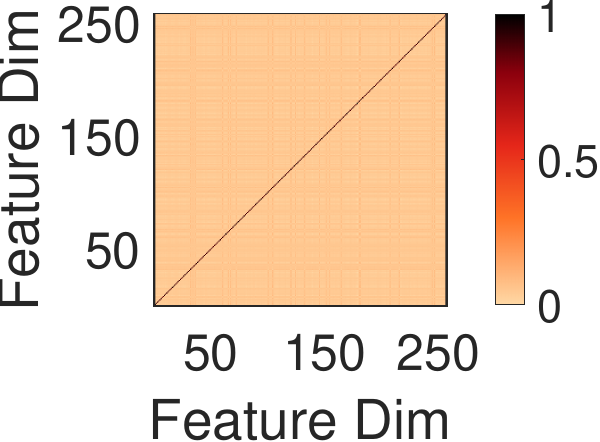}
		\label{fig:corr}
	}\hspace{2pt}
	\subfigure[{\small The power and smoothed result of $\mathbf{h}_{mm}$.}]{
		\centering
		\includegraphics[width=0.18\linewidth]{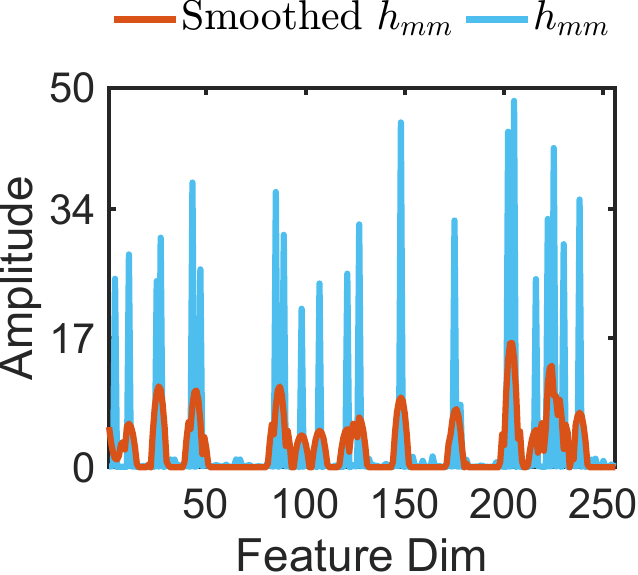}
		\label{fig:p1}
	}\hspace{2pt}
	\subfigure[{\small The power and smoothed result of $\mathbf{h}_{ac}$.}]{
		\centering
		\includegraphics[width=0.178\linewidth]{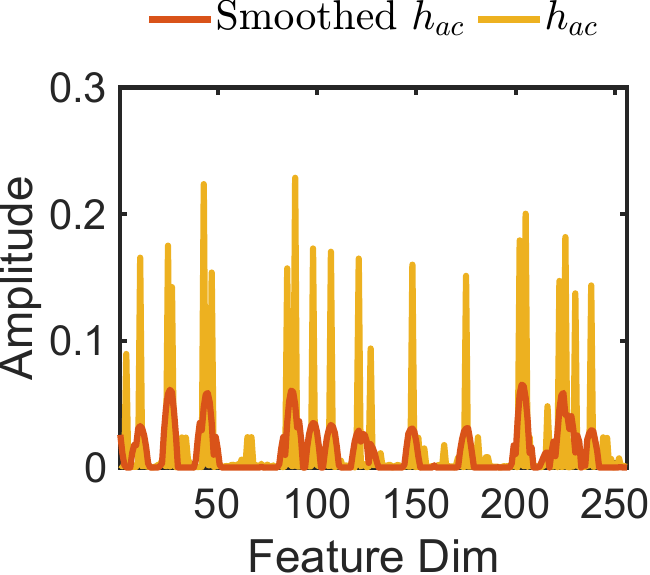}
		\label{fig:p2}
	}\hspace{2pt}
	\subfigure[{\small The power and smoothed result of $\tilde{\mathbf{h}}_{mm}$.}]{
		\centering
		\includegraphics[width=0.18\linewidth]{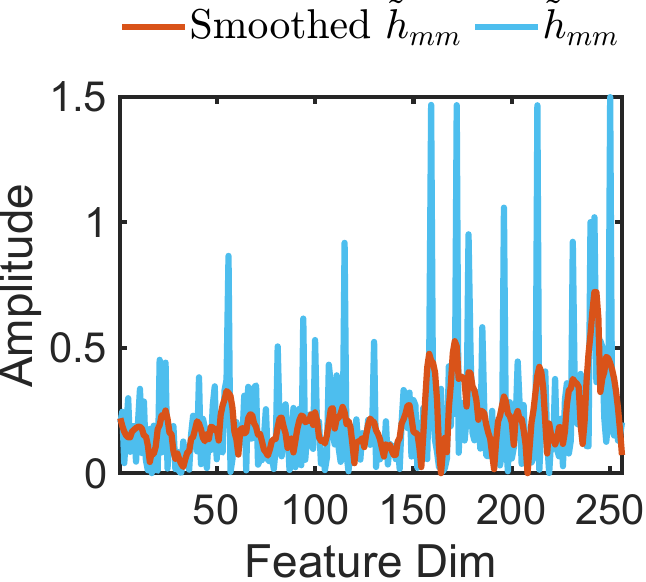}
		\label{fig:p3}
	}\hspace{2pt}
	\subfigure[{\small The power and smoothed result of $\tilde{\mathbf{h}}_{ac}$.}]{
		\centering
		\includegraphics[width=0.18\linewidth]{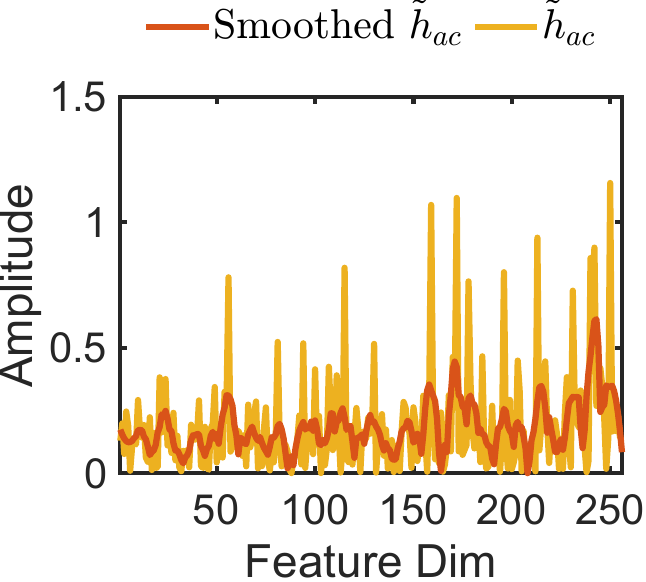}
		\label{fig:p4}
	}
	\vspace{-15pt} 
	\caption{(a) Correlation analysis between radar ($\mathbf{h}_{mm}$) and acoustic ($\mathbf{h}_{ac}$) features. 
		(b--c) Power distribution with smoothed versions of raw $\mathbf{h}_{mm}$ and $\mathbf{h}_{ac}$. 
		(d--e) Scaled features after attention-based scaling ($\tilde{\mathbf{h}}_{mm}$, $\tilde{\mathbf{h}}_{ac}$). }
	\label{fig:scale}
	\vspace{-10pt}
\end{figure*}

\sssec{Feature Alignment.} The mmWave and acoustic modalities capture complementary information but inherently differ in scale, distribution, and semantic structure. To resolve such mathematically incompatibility, \sysname employs alignment heads $\Phi_L(\cdot)$ implemented by linear projector to map features into a unified semantic space. The alignment is achieved with  Cross-modal Barlow Twins objective~\cite{btloss} (see Appendix~\ref{app:btloss} for details), which enforces dimension-wise correspondence by minimizing the difference between the cross-correlation matrix and the identity matrix. As a result, the projected representations $\mathbf{h}=\Phi_L(x)$ from both modalities are aligned in a shared semantic coordinate system, establishing a valid mathematical basis for subsequent disentanglement. Fig.~\ref{fig:corr} visualizes the strong cross-modal correlations, validating the effectiveness of this alignment.

\sssec{Dimension-wise Scaling.} 
\label{sssec:scaling}
While the alignment heads effectively unify the feature dimensionality, they fail to account for the power disparities due to distinct sensing mechanisms. As shown in Fig.~\ref{fig:p1}-\ref{fig:p2}, we visualize the distribution trends across feature dimensions in both modalities. The filtering method is detailed in Appendix.~\ref{app:sgfiltering}. The results indicate that the aligned mmWave and acoustic features exhibit highly similar distributions but significant magnitude discrepancies (i.e., energy  mismatch), preventing effective removal.  

To ensure mathematically valid subtraction, we introduce a cross-modal attention–based calibration module. Unlike rigid global normalization, this module utilizes attention weights to perform adaptive, dimension-wise power scaling, thus allowing for fine-grained calibration of cross-modal power mismatches (approximate mean magnitude in Fig.~\ref{fig:p3}-\ref{fig:p4}) while strictly preserving the semantic structure established during alignment (as the similarity comparison shown in Fig.~\ref{fig:scale}). Consequently, we obtain latent representations in the shared space that are \textbf{semantically aligned, distribution-consistent, and comparable in scale}:

\begin{equation}
\begin{aligned}
\tilde{\mathbf{h}}_{mm} &= Attn_{mm}(\mathbf{h}_{mm},\mathbf{h}_{ac})\odot\mathbf{h}_{mm},  \\
\tilde{\mathbf{h}}_{ac} &= Attn_{ac}(\mathbf{h}_{ac},\mathbf{h}_{mm})\odot\mathbf{h}_{ac}
\end{aligned}
\label{eq:power_scale} \nonumber
\end{equation}
where $\tilde{\mathbf{h}}$ denotes the aligned and calibrated feature, and $\odot$ is the Hadamard product. The dimension-wise scaling moudle $Attn_{mm}$ and $Attn_{ac}$ are implemented via cross-modal attention~\cite{crossattention} to adaptively rescale each feature dimension, enabling fine-grained calibration of inter-modal power imbalance.
As shown in Fig.~\ref{fig:p3}-~\ref{fig:p4}, our scaling  equalizes the energy across feature dimensions, revealing that effective information is distributed across all dimensions.

\sssec{Subtractive Removal.} After scaling, we then derive the material embedding $\mathbf{h}^\mathcal{M}$ by subtracting the aligned acoustic feature from the aligned mmWave feature (i.e., the material-centric feature $\mathbf{h}_{mc} = \tilde{\mathbf{h}}_{\text{mm}} - \tilde{\mathbf{h}}_{\text{ac}}$):
\begin{equation}
\mathbf{h}^\mathcal{M} = \Phi_\mathcal{M}(\tilde{\mathbf{h}}_{\text{mm}} - \tilde{\mathbf{h}}_{\text{ac}})
\label{eq:subtraction_feature}
\end{equation}
where $\Phi_\mathcal{M}(\cdot)$ is a lightweight material head, mapping the material-centric feature into a latent material subspace. Similarly, geometry embedding is defined as:
\begin{equation}
\mathbf{h}^\mathcal{G} = \Phi_{\mathcal{G}}(\tilde{\mathbf{h}}_{\text{mm}} +  \tilde{\mathbf{h}}_{\text{ac}})
\label{eq:spatial_feature}
\end{equation}
which aggregates geometry information from both modalities (i.e., the geometric-centric feature $\mathbf{h}_{gc} = \tilde{\mathbf{h}}_{\text{mm}} + \tilde{\mathbf{h}}_{\text{ac}}$). This representation is used to regularize the disentanglement process by explicitly modeling the geometric factors shared across sensing modalities. Preliminary experiments in Sec.~\ref{sec:DominantShared GeometryVerification} imply that the intersection of the two feature spaces corresponds to geometry.  By executing the subtraction in Eq.~\ref{eq:subtraction_feature}, we effectively suppress shared geometric context while preserving material information. 

To further enforce the separation between material  embedding and geometry  embedding, we introduce two complementary regularization terms. First, we introduce a \textbf{feature orthogonality constraint} $\mathcal{L}_{\mathrm{corr}}$ (see Appendix~\ref{app:corrloss} for details) to minimize statistical correlation between $\mathbf{h}^\mathcal{M}$ and $\mathbf{h}^\mathcal{G}$. 
Minimizing $\mathcal{L}_{\mathrm{corr}}$ explicitly penalizes cross-correlation between the two representations, enforcing orthogonality between the material and geometry subspaces. As a result, material embedding are
encouraged to be statistically independent of geometry embedding, improving separation in the learned embedding space.

Second, we employ a \textbf{reconstruction loss} that reconstructs the aligned representations from the disentangled features to prevent the disentangled geometry features from collapsing to noise due to information loss:
\begin{align}
\mathcal{L}_{rec}
=
\left\|
\Phi_{rec}\!\left(
\mathbf{h}^\mathcal{M} + \mathbf{h}^\mathcal{G}
\right)
-
\mathbf{h}_{gc}
\right\|_2^2 \nonumber
\label{eq:reconstruct_feature} 
\end{align}
which enforces that the material and geometry subspaces jointly retain the full information of the original features, rather than collapsing to a degenerate solution. This reconstruction loss ensures $\mathbf{h}^\mathcal{G}$ is informative, since random noise can also be orthogonal to $\mathbf{h}^\mathcal{M}$. Through this disentanglement framework corresponding to the physical sensing model, the module effectively isolates material embedding by suppressing geometry-induced interference.

\subsection{Correcting Residual Cross-Modal Misalignment}
\label{sec:contra}

Although the disentanglement module removes shared geometric structures, the extracted material embedding  $\mathbf{h}^\mathcal{M}$ still suffers from \textbf{residual cross-modal waveform misalignment}. This issue primarily stems from the  differences between the sensing
mechanisms $G^{mm}_{\mathrm{sys}}$ and $G^{ac}_{\mathrm{sys}}$, which cause each modality to perceive the environment in different ways. Specifically, mmWave beamforming emphasizes spatially localized reflections, whereas quasi-spherical acoustic propagation integrates responses over a broader region. 
This makes simple cross-modal subtraction insufficient for fully canceling such modality-specific effects. Moreover, after subtraction, the material signatures are weak and often comparable in magnitude to the residual components, complicating clean extraction of material representations.

\sssec{Contrastive Learning for Misalignment Correction.} 
To mitigate residual cross-modal misalignment, \sysname leverages a key intuition: intrinsic material properties are position-invariant, whereas residual noise varies spatially. As a result, misalignment tends to disperse features of the same material in the embedding space. We therefore adopt an  \textbf{inter-sample contrastive learning} strategy
that clusters representations of the same material across different locations while separating distinct materials. This performs implicit multi-sample averaging, suppressing spatially varying noise and improving robustness to waveform misalignment.
\begin{figure}[t]
    \centering
    \includegraphics[width=0.65\linewidth]{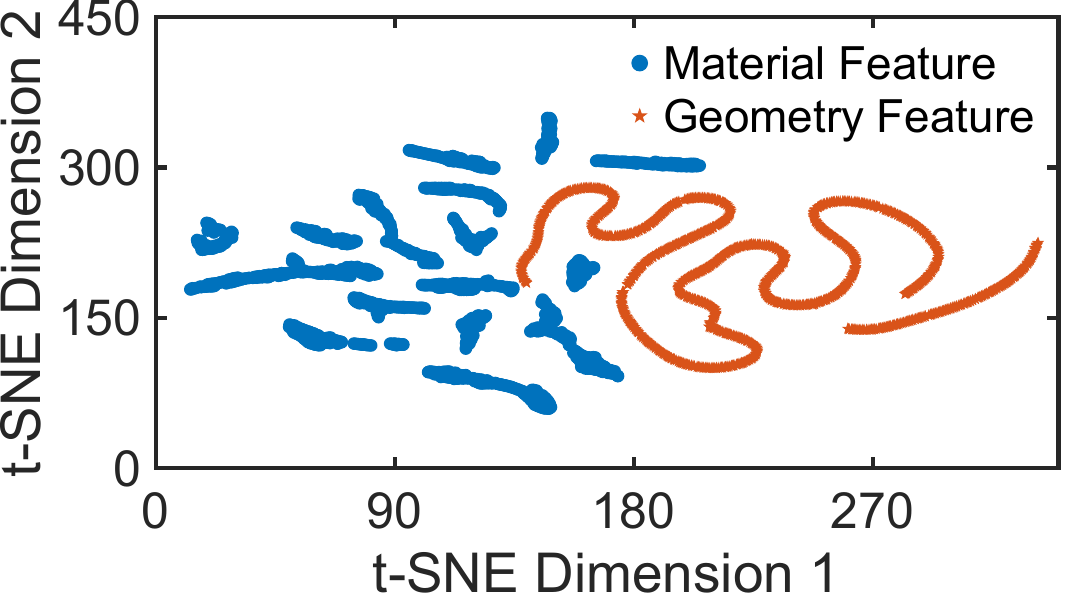}
	\vspace{-10pt}
    \caption{T-SNE visualization of disentangled features, where material and geometry features are separated.
	}
	\vspace{-15pt}
    \label{fig:tsne}
\end{figure}
In practice, positive pairs are samples sharing the same material label but collected at different locations. Notably, the dataset \textbf{does not require precise metric coordinates}. Formally, we minimize the InfoNCE loss~\cite{infoNCE} (Appendix~\ref{app:infonce}) $\mathcal{L}_{con}$ suppresses residual in the material feature subspace, yielding robust material representations. The reconstruction loss preserves the full input information, which inevitably encodes Gaussian system noise. After removing dominant geometric components, the residual intrinsic material signatures are weak and comparable to remaining noise. The contrastive learning mechanism naturally mitigates this noise through implicit multi-sample averaging, thereby enhancing robustness. Finally, Fig.~\ref{fig:tsne} presents the t-SNE visualization of features after subtractive multimodal disentanglement and contrastive learning correction. The plot shows clear separation of material and geometry embedding, confirming successful material representation extraction.

\subsection{Material Classification}
\label{sec:cls}
We employ a classification head to categorize the disentangled and denoised material features using cross-entropy loss $\mathcal{L}_{cls}$~\cite{crossentropy}. The model is optimized using a composite loss function combining multiple objectives:
\begin{equation}
\mathcal{L}_{\text{total}} =\mathcal{L}_{\text{cls}} +
\lambda_1 \mathcal{L}_{\text{corr}} + \lambda_2 \mathcal{L}_{\text{rec}} + \lambda_3 \mathcal{L}_{\text{con}}
\end{equation}
where $\mathcal{L}_{cls}$ implement the cross-entropy classification loss, $\mathcal{L}_{corr}$ enforces feature decorrelation, $\mathcal{L}_{rec}$ measures reconstruction quality, and $\mathcal{L}_{con}$ denotes contrastive learning. The weighting coefficients $\lambda_1=1$, $\lambda_2=0.01$, and $\lambda_3=0.01$, were empirically determined to balance the loss components. 

\section{Cross-device Generalization}
\label{sec:general}
\sssec{Cross-device Distribution Shift.} In real-world deployments, \sysname is expected to generalize across heterogeneous sensing devices with minimal adaptation effort. However, variations in mmWave and acoustic hardware specifications $G_{sys}$ (e.g., transmit power, sensitivity, and frequency response) can induce non-negligible shifts in the learned feature distributions. Re-collecting a large-scale dataset for every new device is impractical, while the limited amount of target-device data further makes such hardware-induced shifts difficult to model and correct after deployment.
\begin{figure}[t]
	\centering
	\includegraphics[width=1\columnwidth]{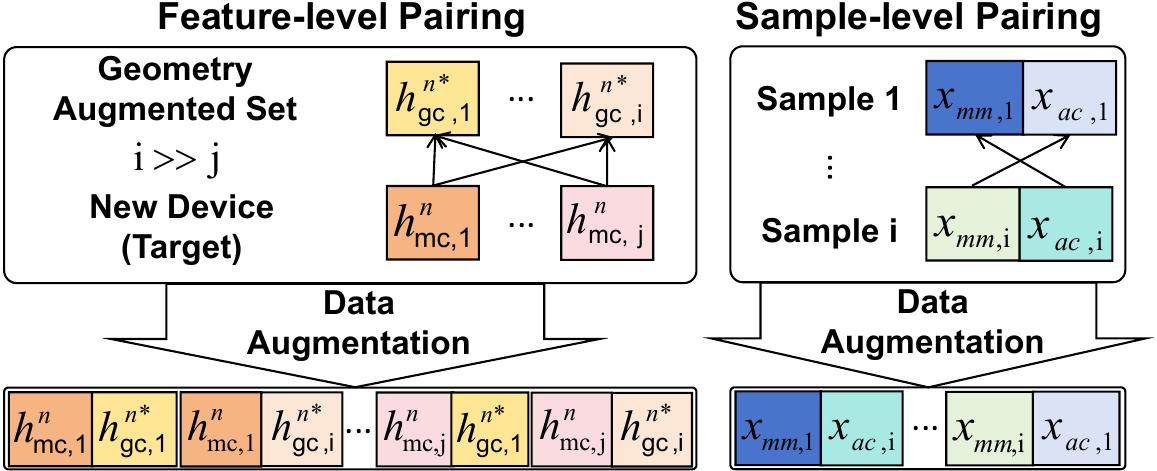}
	\vspace{-20pt}
	\caption{Paired-data adaptation enables efficient learning of cross-device distribution shifts with limited new-device data.}
	\vspace{-10pt}
	\label{fig:pairdata}
\end{figure}
\sssec{Pairing-based Cross-device Adaptation.} To address this challenge, we propose a pairing-based adaptation strategy to expand training diversity under limited target-device data, tackling two key deficiencies. First, only a small number of multimodal samples are available, which limits the diversity of cross-modal patterns that can be observed during adaptation. We therefore adopt \textbf{sample-level pairing}, which recombines mmWave and acoustic samples from different captures to augment the effective training set and expose the model to richer sample variations. Second, the collected target-device data often cover only a few geometric conditions, resulting in insufficient geometry diversity for robust adaptation. To compensate for this limitation, we adopt \textbf{feature-level pairing}, which leverages geometry-centric variations extracted from abundant source-device data to enrich the target-device feature space. These two pairing strategies are enabled by different structural properties of \sysname: sample-level pairing relies on multimodal observations, whereas feature-level pairing relies on the disentangled material- and geometry-centric representations.

Specifically, we consider a \textit{one-site calibration} setting, where each material on a new device is calibrated using a few seconds of data collected once at a single location (i.e., without strictly constraining the exact geometric relationship during calibration). Such a short collection window naturally contains multiple samples, and no additional relocation is required for that material. Under this setting, \textbf{sample-level pairing} directly augments the target-device data by recombining mmWave and acoustic samples from different captures that share the same material label and geometric condition, thereby forming additional multimodal pairs. In contrast, \textbf{feature-level pairing} compensates for the limited geometric coverage of one-site calibration. For geometry-centric features in old device $\mathbf{h}^{o}_{gc}$, we compute geometry differences as $\Delta \mathbf{h}^{o}_{gc} = \mathbf{h}^{o}_{gc,i} - \mathbf{h}^{o}_{gc,j}$, where $i$ and $j$ denote random samples that share the same material label but differ in geometry. We then transfer these geometry-centric variations to the target-device features by constructing $\mathbf{h}_{gc}^{n*} = \mathbf{h}_{gc}^{n} + \Delta \mathbf{h}^{o}_{gc}$, and pair the augmented geometry-centric features $\mathbf{h}_{gc}^{n*}$ with the target-device material-centric features $\mathbf{h}_{mc}^{n}$ to construct an augmented set of trainable feature pairs, thereby enriching the geometry diversity available for target-device adaptation. Together, these two schemes complement each other by increasing sample diversity and geometry diversity in the target-device data, enabling robust adaptation.


\vspace{-5pt}
\section{Evaluation}

\vspace{-5pt}
\subsection{Evaluation Methodology}\label{sec:evaluation_methodology}

\begin{figure}[t]
	\centering
	\subfigure[{\small \sysname.}]{%
		\centering
		\includegraphics[width=0.235\linewidth]{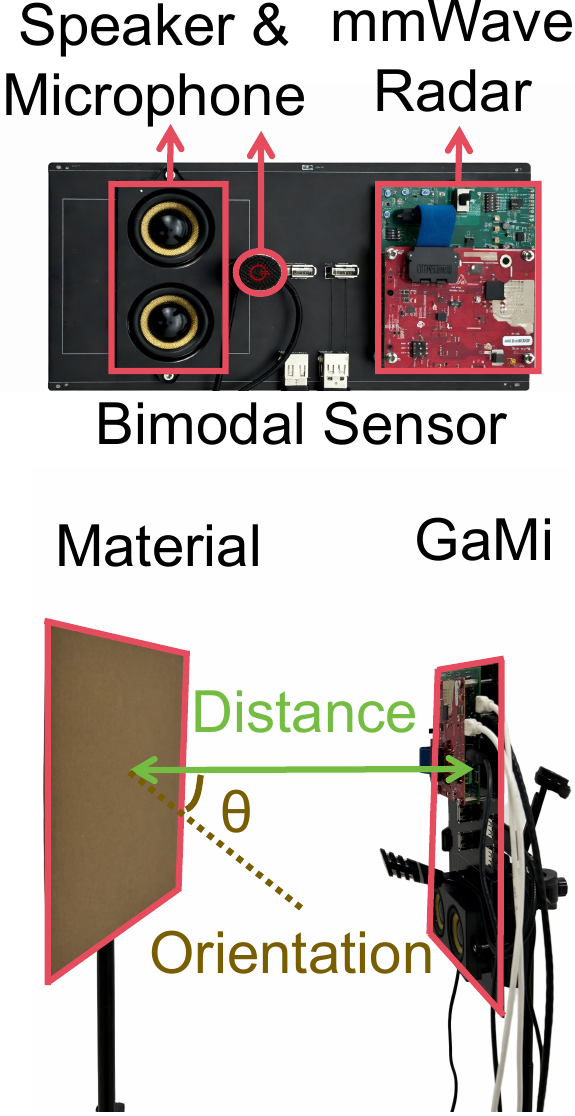}%
		\label{fig:gami}%
	}\hfill
	\subfigure[{\small Experiment materials.}]{%
		\centering
		\includegraphics[width=0.765\linewidth]{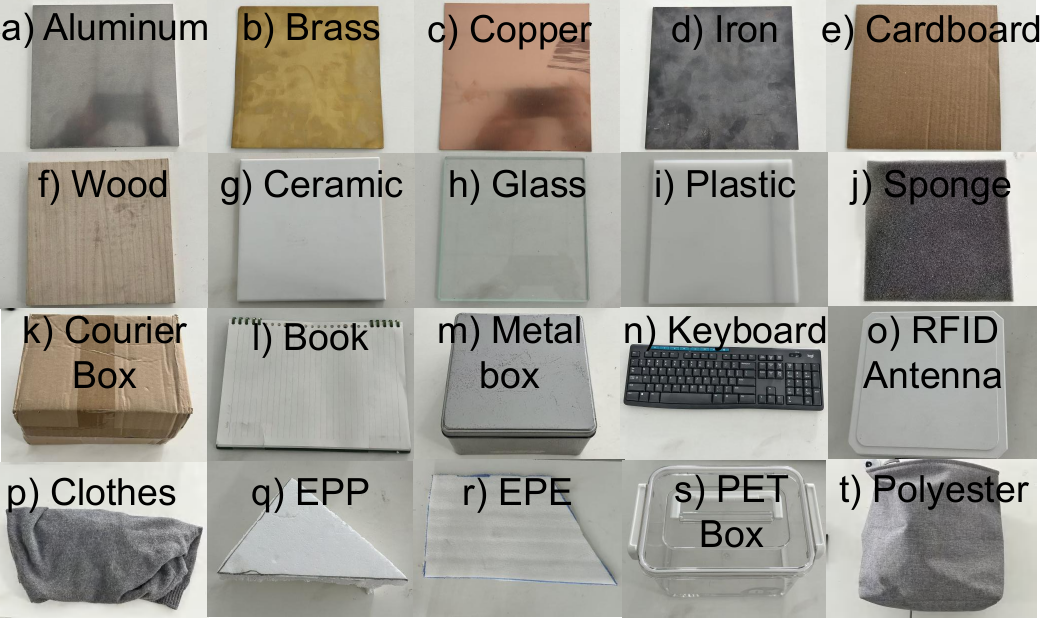}%
		\label{fig:material}%
	}
	\vspace{-15pt}
	\caption{Experimental setup of \sysname.}
	\vspace{-15pt}
    \label{fig:material_gami}
\end{figure}

\noindent\textbf{Implementation.}
\sysname is built from off-the-shelf mmWave and acoustic components mounted on a simple PCB board, thereby forming the bimodal sensor shown in Fig.~\ref{fig:gami}. The mmWave front-end uses a TI IWR1843 transceiver~\cite{TI_IWR1843} operating at 76--81$GHz$, together with a DCA1000EVM~\cite{TI_DCA1000EVM} for data capture. The acoustic front-end uses a low-cost loudspeaker~\cite{firefly_speaker} to emit ultrasonic Zadoff--Chu pulse sequences over 17--22$kHz$, while the reflected signals are recorded at 48$kHz$ by a microphone module~\cite{newmine_zm12}. 
We implement synchronized multimodal data acquisition in MATLAB 2024b, extracting Range--Azimuth heatmaps and 2D-AoA spectra for mmWave, and CIRs for acoustics. Model training and inference are implemented on a server with a RTX 4090 GPU (24$GB$ memory).

\noindent\textbf{Material Selection.}  Comprehensive evaluation is conducted  on 20 common materials (Fig.~\ref{fig:material}), covering a wide range of acoustic and RF reflection characteristics. The selection covers three criteria: (1) prevalence in daily environment; (2) heterogeneous materials such as solid-rigid ('e' cardboard) and hollow-flexible materials ('k' courier box); (3) compositional complexity, distinguishing uniform substances ('a'-'e' pure metallic) from composites ('m' metal box). 

\noindent\textbf{Dataset Preparation.} 
We collected data by positioning mmWave radar and acoustic sensors at 0.5--1.4$m$, covering orientations 0$^\circ$–30$^\circ$ relative to the surface normal (Fig.~\ref{fig:gami}). Specifically, we sampled 10 distances at 10$cm$ intervals, with 3 orientations per distance, totaling 30 locations. Notably, the process does not require explicit spatial coordinates; we only record discrete location indices. Beyond 1.4$m$, acoustic signals are too weak; using higher-power devices could extend the range.

\noindent\textbf{Baselines and Metrics.} We compare \sysname against two representative  baselines: \textit{1) Acoustic-only}: To evaluate how \sysname resolves ambiguities, we construct this baseline by ablating the mmWave branch from \sysname, retaining only acoustic sensing and classification.  \textit{2) MID}~\cite{mid}: MID is an mmWave-only method that leverages domain adaptation for distance and orientation robustness. We compare \sysname with MID to highlight our geometry-agnostic capabilities, particularly in cases where MID's performance is limited by variations in surface geometry. We adopt the classification accuracy as the primary performance metric.

\noindent\textbf{Leakage-Free Split Protocol.} To avoid hidden leakage in geometric generalization experiments, we enforce disjoint train/test splits at three levels:
\textit{1) Overall-geometry split.} All distance--orientation combinations are pooled and randomly split, so a test sample may share its distance or orientation with training samples, but not the exact combination.
\textit{2) Unseen-geometry split.} For unseen-distance, unseen-orientation, and unseen-shape evaluations, we enforce strict disjoint splits over the tested factor, i.e., any test distance bin, orientation bin, or shape category is entirely excluded from training.
\textit{3) Object-instance split.} All samples from the same physical object instance (same item ID) are placed entirely in either train or test, whenever a material class contains multiple objects (e.g., the application case studies).

\vspace{-5pt}
\subsection{Overall Performance}
\begin{figure}[t]
	\centering
    \vspace{-5pt}
	\subfigure[\sysname.]{%
		\includegraphics[width=0.33\linewidth]{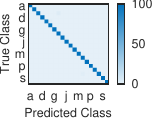}%
		\label{fig:ours}%
	}\hfill
	\subfigure[Acoustic-only.]{%
		\includegraphics[width=0.33\linewidth]{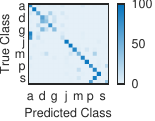}%
		\label{fig:ac2}%
	}\hfill
	\subfigure[MID~\cite{mid}.]{%
		\includegraphics[width=0.33\linewidth]{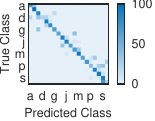}%
		\label{fig:mid}%
	}
	\vspace{-10pt}
	\caption{Confusion matrices of overall classification. \sysname achieves 95.2\% accuracy, outperforming acoustic-only (62.05\%) and MID (71.55\%) baselines.}
	\label{fig:confusion_matrices}
	\vspace{-13pt}
\end{figure}

\subsubsection{Overall Classification} We evaluate \sysname across all 30 configurations using 5-fold cross-validation following the \textit{overall-geometry split} protocol and report the average classification accuracy. As shown in Fig.~\ref{fig:confusion_matrices}, \sysname achieves an average accuracy of 95.2\%, significantly outperforming the Acoustic-only (62.05\%) and MID (71.55\%) baselines. The performance gap highlights the limitations of unimodal sensing: the baselines exhibit confusion among materials with similar physical properties (e.g. glass, aluminum, and polymers) due to similar acoustic impedances or dielectric constants. By contrast, \sysname effectively disambiguates these categories by fusing complementary acoustic and RF signatures, providing stable and robust identification across diverse geometries and spatial configurations.

\subsubsection{Unseen Geometry.}\label{sec:unseen} We evaluate the geometry-agnostic capability of \sysname under strictly disjoint train/test splits following the \textit{unseen-geometry split} protocol. We also compare \sysname with the Acoustic-only and MID baselines.

\noindent\textbf{Unseen Distances.} We first independently assess distance independence of \sysname. Specifically, we partitioned the entire dataset by distance, in which 8 distances for training and remaining 2 unseen distances for testing (5-fold cross-validation with random split). As shown in Fig.~\ref{fig:unsee}, \sysname achieves high accuracy (92.21\%), while the MID method yields inferior performance (60.35\%). This decline, compared to MID's overall performance, occurs because the random splits in the overall evaluation allowed the model to implicitly encounter all distance points during training, but now the distances are strictly isolated. In comparison, the Acoustic-only baseline drops to 17.57\%, failing to cope with distance-induced acoustic reflection variations.

\noindent\textbf{Unseen Orientation.} We partitioned the dataset by orientation: 2 orientations (20 groups) for training and 1 orientation (10 groups) for testing (3-fold cross-validation). \sysname achieves 88.72\% accuracy, MID drops to 57.01\% (Fig.~14), and the Acoustic-only baseline performs worst at 15.83\%.

\noindent\textbf{Unseen Shapes.}  To evaluate generalization to unseen shapes, we tested the model directly on materials of entirely unseen shapes. This evaluation encompasses four materials (cardboard, wood, plastic boards, and clothing) across various shapes (circle, triangle, rectangle, and free-form (clothing) targets), where our training mainly based on square shapes. Fig.~\ref{fig:unsee} shows that \sysname achieves the highest accuracy of 89.31\% whereas the baselines degrade sharply: MID (35.23\%) and Acoustic-only (27.28\%). This is because MID is more sensitive to surface geometry than distance and orientation. In contrast, \sysname successfully disentangles material-specific features by subtracting the shared geometries, demonstrating shape-agnostic identification.

\vspace{-5pt}
\subsection{Cross-device Generalization} 
\label{sec:newdevice}
\sysname illustrates cross-device generalization by leveraging the proposed pairing-based adaptation strategy. 
To evaluate this capability, we replace the original mmWave and acoustic sensors with new units of the same hardware models, thereby introducing device-induced distribution shifts. 
For adaptation, we collect 5$s$ of calibration data per material at either one location (one-site calibration) or two locations (two-site calibration), yielding 50 samples per location.
This lightweight adaptation method requires only brief measurements at one or two roughly chosen sites, without precise calibration-point selection for each material, keeping calibration effort minimal.
We then evaluate the adapted model on additional locations at least 0.4$m$ away from the calibration site(s) to ensure sufficient geometric variation.

\begin{figure}[t]
	\begin{minipage}[t]{0.45\linewidth}
		\centering
		\includegraphics[width=\linewidth]{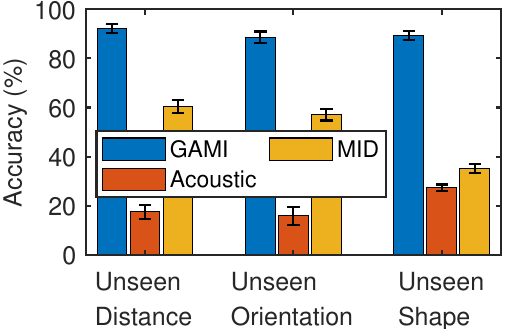}
		\vspace{-15pt}
		\caption{Accuracy under unseen geometry.}
		\label{fig:unsee}
        \vspace{-15pt}
	\end{minipage}\hfill
	\begin{minipage}[t]{0.5\linewidth}
		\centering
		{%
			\includegraphics[width=\linewidth]{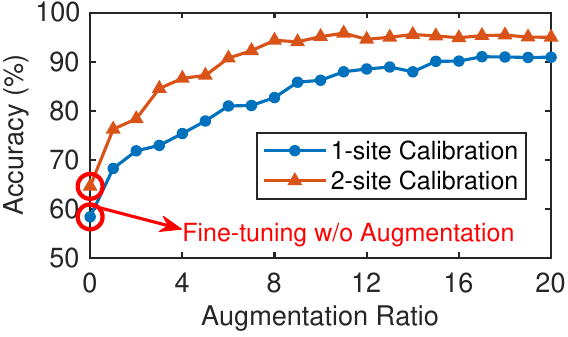}
		}
		\vspace{-15pt}
		\caption{Cross-device generalization.}
		\label{fig:newdevice}
	\end{minipage}
	\vspace{-15pt}
\end{figure}

As shown in Fig.~\ref{fig:newdevice}, the naive fine-tuning without augmentation achieves only $58.37\%$ and $64.58\%$ under one-site calibration and two-site calibration, respectively, highlighting the substantial distribution shift between the original and new sensors.
In comparison, the classification accuracy improves steadily as the augmentation ratio increases, and finally converges to $91.01\%$ and $95.81\%$, indicating that the paired data provide additional multimodal combinations and geometry variations that are absent from the limited calibration samples. Interestingly, the slightly higher 2-site performance than that in Sec.~\ref{sec:unseen} is likely because, unlike the strictly \textit{unseen-geometry split}, the source-domain data in this experiment may already cover part of the tested geometric conditions.
These results show that pairing-based adaptation effectively mitigates cross-device distribution shifts with adequate human effort, making \sysname more practical for real-world deployment.

\vspace{-5pt}
\subsection{Impact of Factors}
\begin{figure*}[t]  
	\centering
	\subfigure[{\small Impact of distance.}]{%
		\centering
		\includegraphics[width=0.19\linewidth]{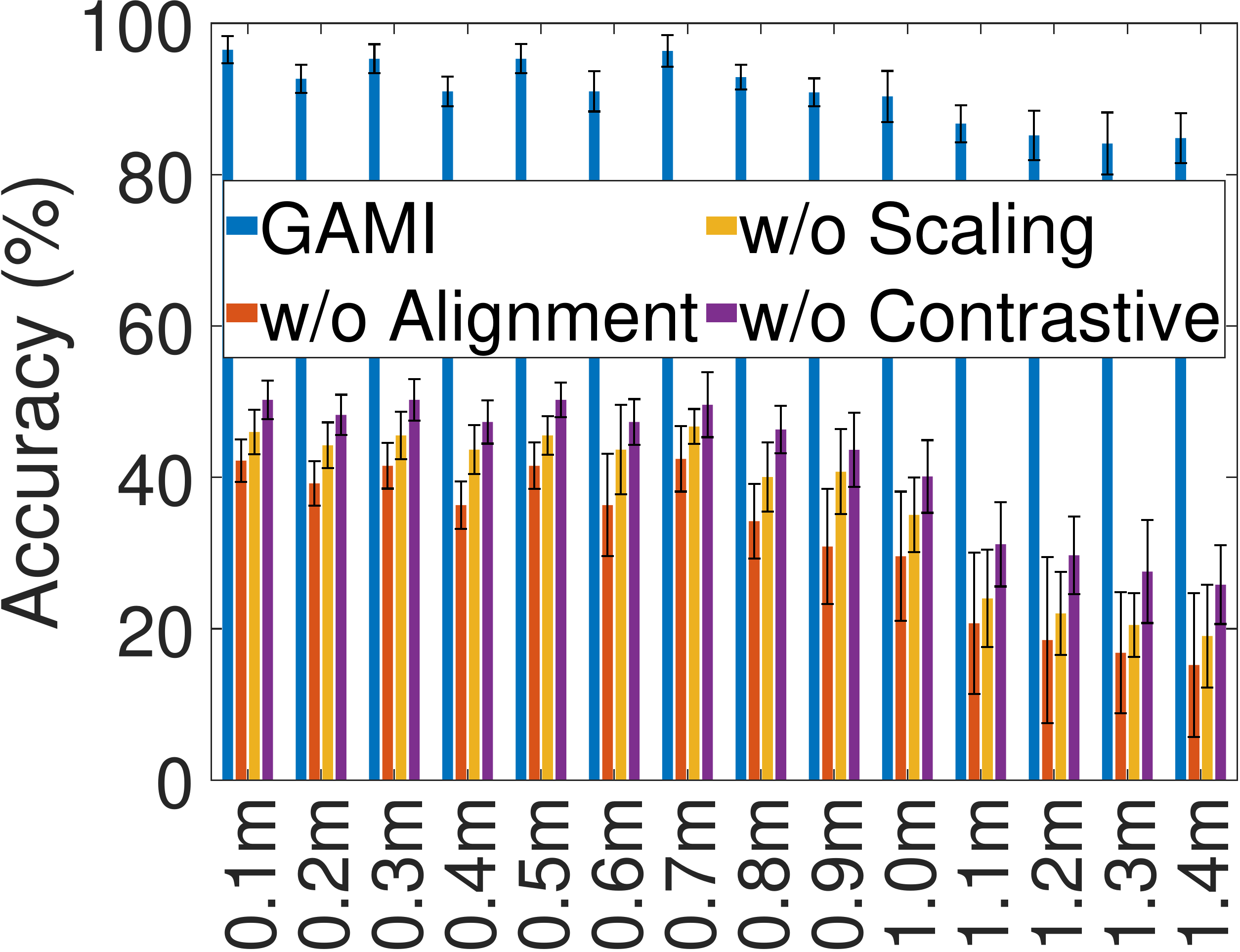}%
		\label{fig:imdis}%
	}\hfill
	\subfigure[{\small Impact of orientation.}]{%
		\centering
		\includegraphics[width=0.19\linewidth]{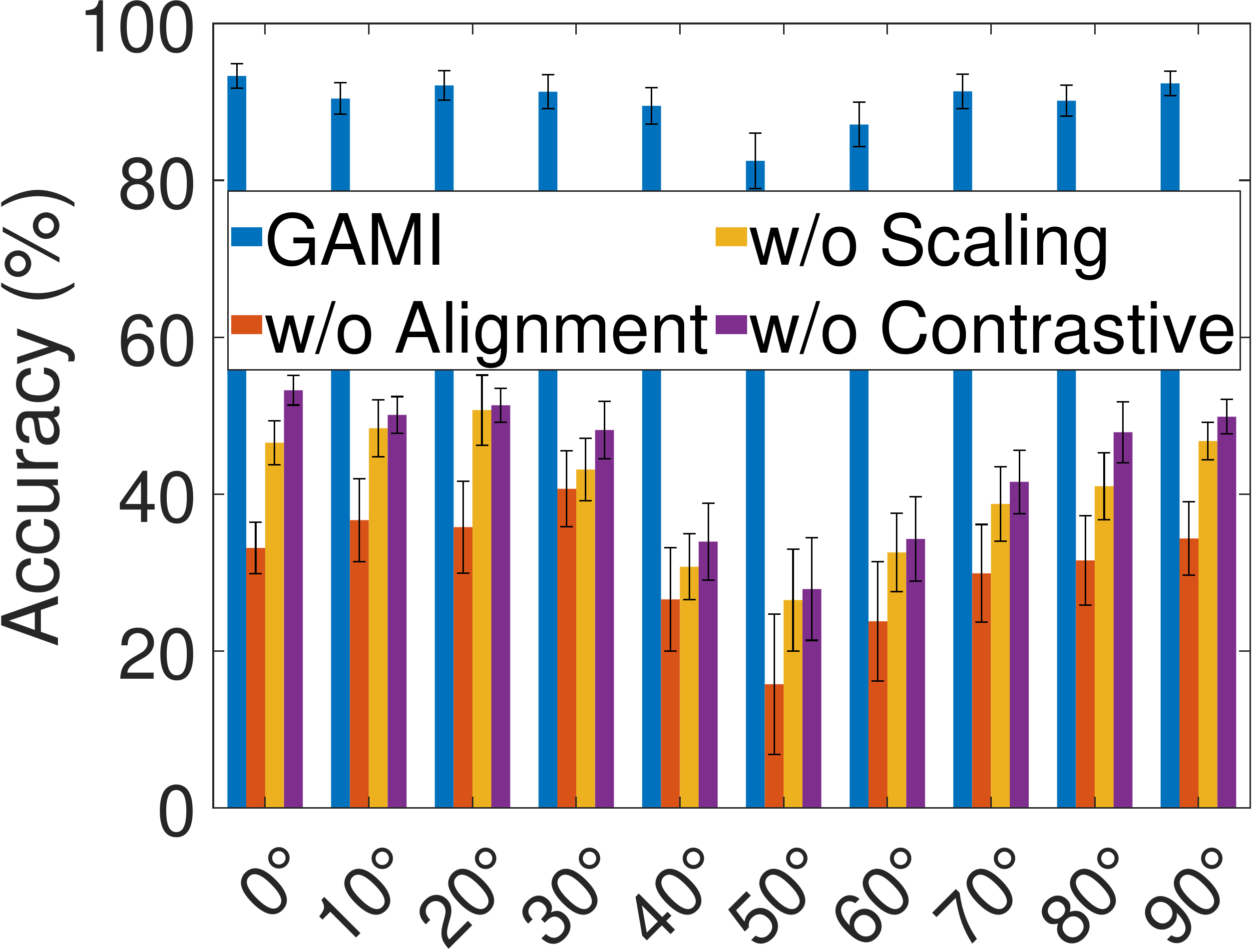}%
		\label{fig:imangle}%
	}\hfill
	\subfigure[{\small Impact of shape.}]{%
		\centering
		\includegraphics[width=0.19\linewidth]{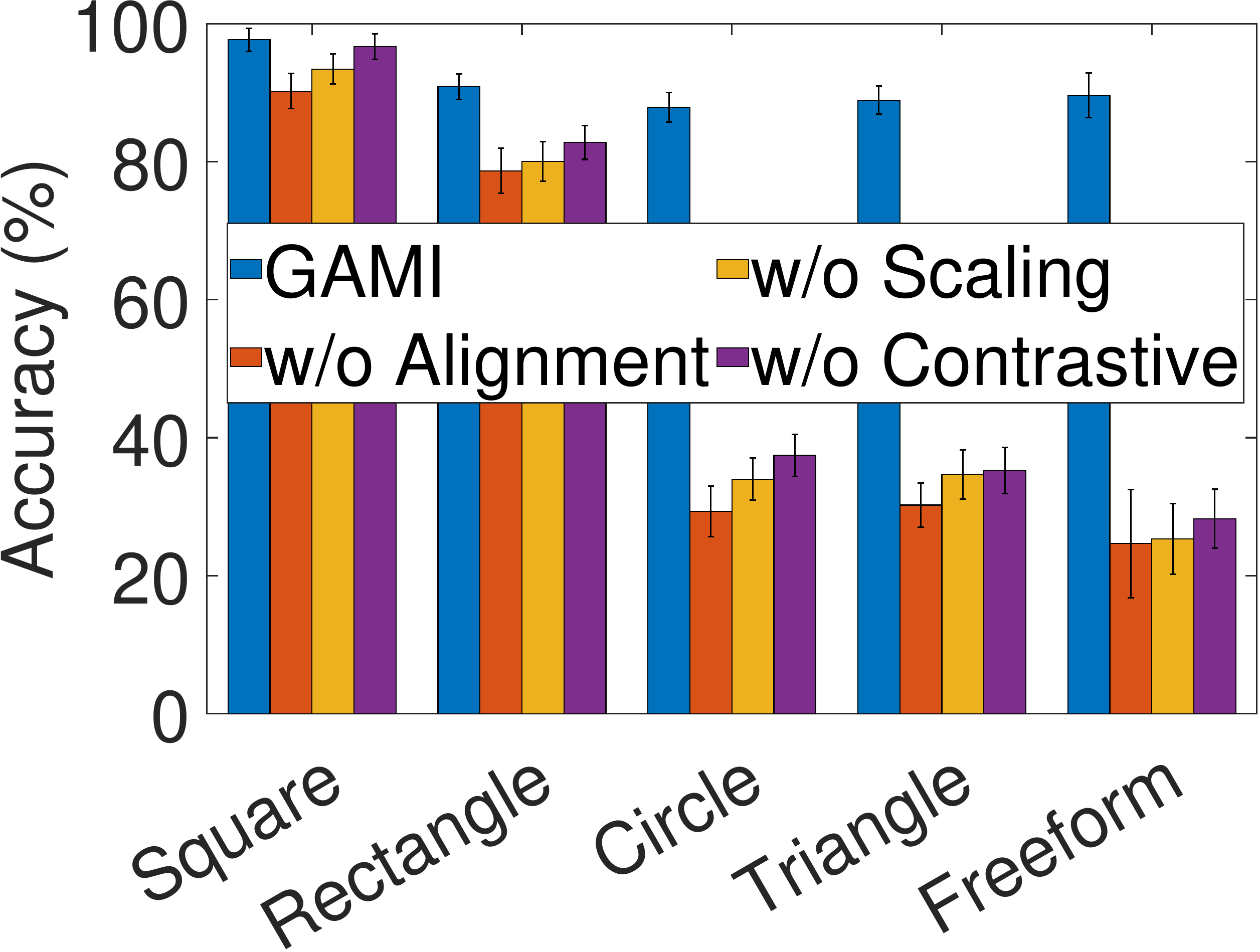}%
		\label{fig:imshape}%
	}\hfill
	\subfigure[{\small Impact of environment.}]{%
		\centering
		\includegraphics[width=0.19\linewidth]{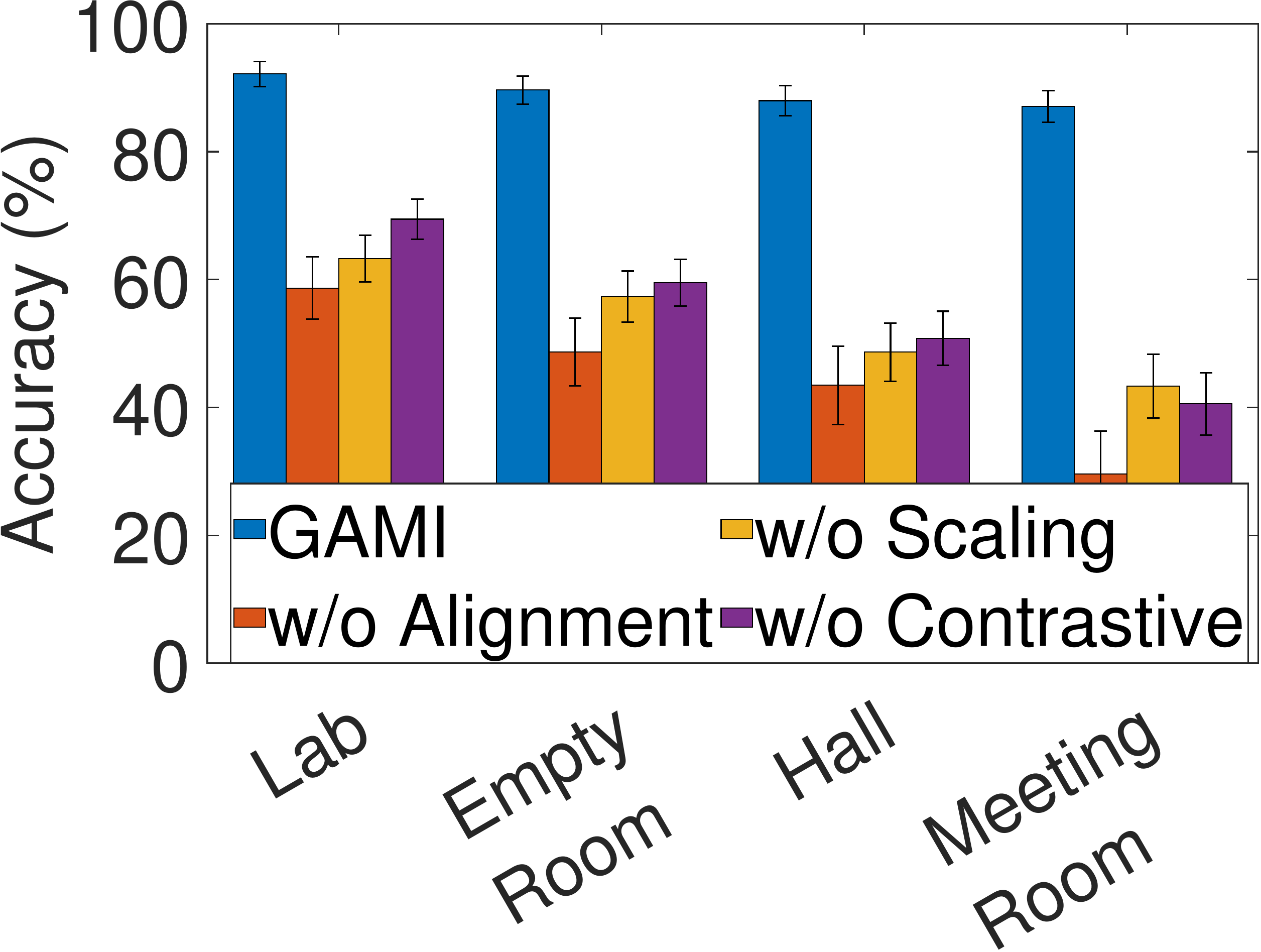}%
		\label{fig:imenv}%
	}\hfill
	\subfigure[{\small Impact of noise.}]{%
		\centering
		\includegraphics[width=0.19\linewidth]{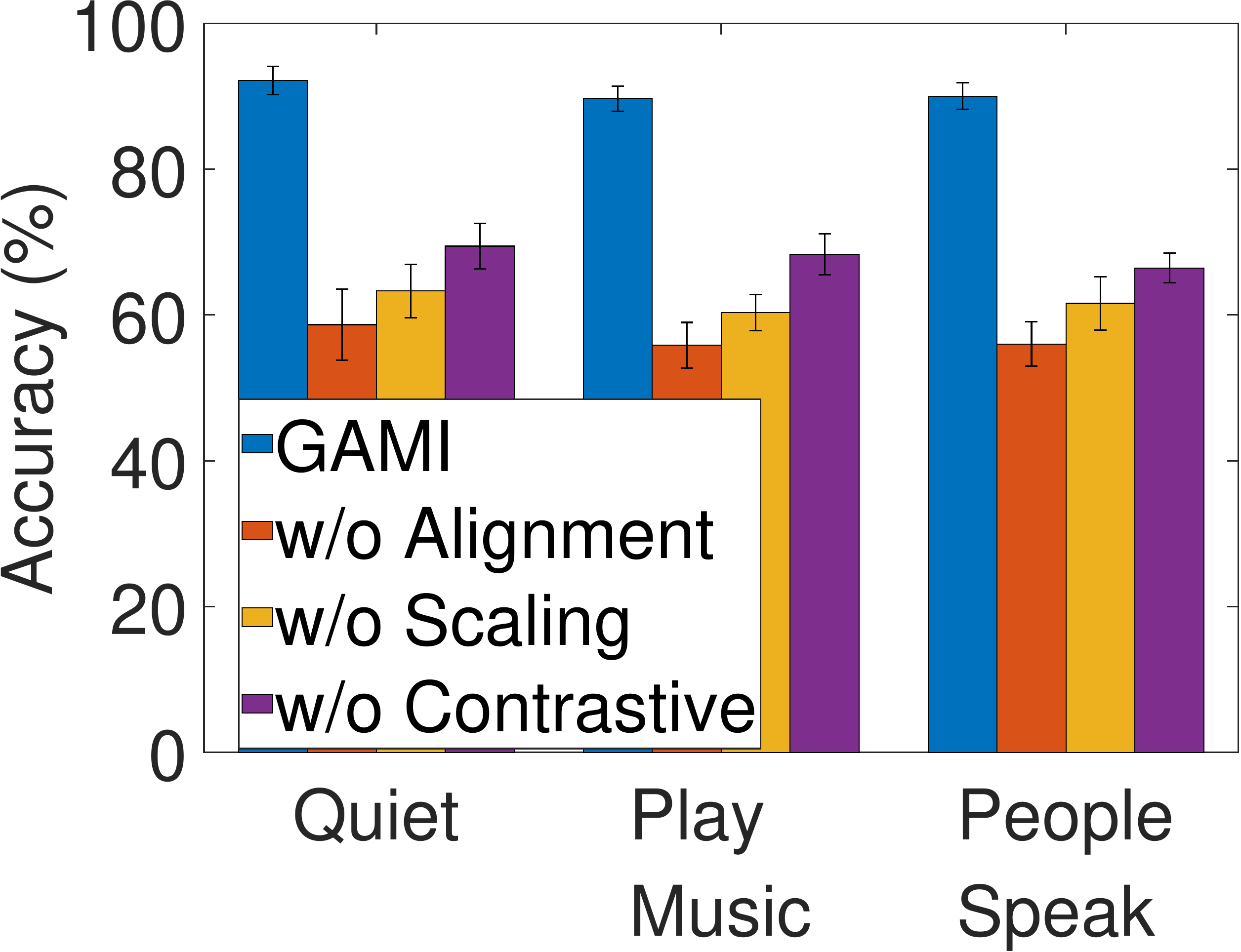}%
		\label{fig:imnoise}%
	}
	\vspace{-12pt}
	\caption{Impact of different factors on \sysname{}’s classification accuracy.
		(a) distance, (b) orientation, (c) shape, (d) environment, and (e) noise. Ablation study is performed to quantify the effectiveness of our design.}
	\vspace{-10pt}
	\label{fig:impact}
\end{figure*}
\sysname uses a cross-modal subtractive framework to achieve robust material identification. In this section, we evaluate the robustness of \sysname in real-world settings and quantify the effectiveness of proposed design choices. Specifically, four models were trained for ablation study: (1) full \sysname design; (2) remove feature alignment module (Sec.~\ref{sec:disent}); (3) remove power scaling module (Sec.~\ref{sec:disent}); and (4) remove contrastive learning module (Sec.~\ref{sec:contra}). 

\noindent\textbf{Impact of Distance.} Sensing distance fundamentally dictates the Signal-to-Noise Ratio (SNR) via signal attenuation. To ensure distance-invariance, we evaluated \sysname across a 0--1.4$m$ range using stratified random sampling, binning results at 0.1$m$ intervals with random perturbations of $\pm5cm$. 
As shown in Fig.~\ref{fig:imdis}, \sysname maintains a robust accuracy of 90.33\% at 1$m$, significantly outperforming its ablated variants. The ablation study underscores the synergy of our proposed modules: the alignment module is most critical (29.58\%), as its absence disrupts the shared embedding required for feature decoupling; power scaling (35.05\%) and contrastive learning (40.12\%) follow, respectively suppressing amplitude-related interference and mitigating residual cross-modal mismatches. Moreover, as the distance exceeds 1$m$, the accuracy of \sysname slightly drops. This degradation is primarily due to the reduced SNR of the acoustic device. 

\noindent\textbf{Impact of Orientation.} Orientation affects the area of the observed object that reflects the signal. To investigate the impact of orientation variation, we fixed the sensing distance at 0.5$m$, and employed stratified random sampling across orientations from $0^\circ$ (frontal view) to $90^\circ$ with random perturbations of $\pm5^\circ$. We selected three box-type objects (Courier Box, Metal Box, and PET Box) for the reason that they can provide relatively stable reflections even at large incident orientation. As illustrated in Fig.~\ref{fig:imangle}, \sysname maintains a robust accuracy of over 82.48\% across all orientations, outperforming all variants. Notably, all models experience a performance dip within the $40^\circ$--$60^\circ$ range, which is primarily driven by edge-effect. 
In this interval, the object edges face the sensors, causing dispersed scattering and lower SNR; however, performance then partially recovers at larger orientations as the side surfaces provide stronger reflections.
\noindent\textbf{Impact of Shape.} Object shape dictates reflecting profiles, as different shapes create distinct wavefront distortions. To evaluate object shape impact, experiments were conducted at a fixed distance of 0.5$m$ and a frontal viewing ($0^\circ$). Similar to the experimental setup in Sec.~\ref{sec:unseen}, we tested multiple unseen shapes and compared with the ablation variants. As illustrated in Fig.~\ref{fig:imshape}, \sysname maintains a robust accuracy of over 87.88\%, whereas removing the power scaling or contrastive learning modules leads to a sharp decline to 25.33\% and 28.25\%, respectively. This degradation highlights that without these modules, shape-induced variations introduce significant cross-modal interference. Notably, rectangular targets exhibit the highest accuracy among unseen shapes, likely due to their structural similarity to the training samples, which offers only small geometric shifts.

\noindent\textbf{Impact of Environment.} Environmental variations introduce multipath interference. To evaluate cross-environment generalization, we trained \sysname using 80\% of data from a lab setting and tested it on the remaining 20\%, as well as on unseen environments including empty room, hall, and furnished meeting room. As shown in Fig.~\ref{fig:imenv}, \sysname maintains a robust accuracy over 87.07\% across all scenarios. Notably, in the furnished meeting room, the variant without contrastive learning underperforms the one without power scaling for the first time. This shift occurs because high environmental complexity amplifies spatially varying residual discrepancies and waveform misalignments between modalities. Without the contrastive module to suppress these residuals, the system becomes susceptible to distortions. Overall, these results demonstrate that \sysname achieves superior environmental invariance, effectively isolating material signatures from surrounding interference. 

\noindent\textbf{Impact of Noise.} To evaluate performance under auditory interference, we tested the model that is trained in a quiet environment in scenarios featuring background music and human speech. As shown in Fig.~\ref{fig:imnoise}, \sysname maintains a robust accuracy of at least 89.67\% across all conditions. This high noise immunity primarily stems from the system’s operation in the ultrasonic frequency band, which inherently filters out audible environmental noise. 

\noindent\textbf{Real-Time Capability.}
\sysname aquires 10 multimodal samples per second, while material inference takes only 0.11$ms$ per sample on an NVIDIA GeForce RTX 4090 GPU, readily supporting real-time deployment. Such detecting rate is sufficient for real-world material identification applications.

\begin{figure}[t]
	\centering
	\subfigure[{\small Orientation-agnostic cup material identification}]{%
		\includegraphics[width=0.4\linewidth]{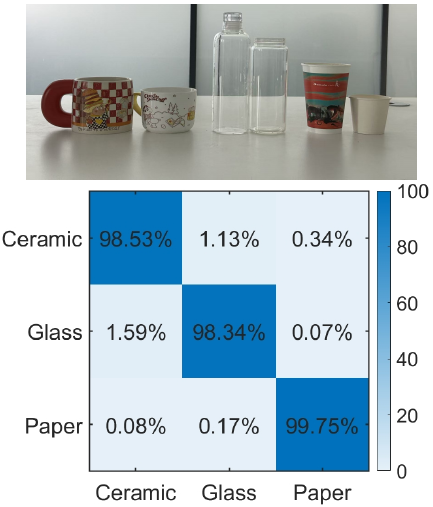}%
		\label{casecup}%
	}\hfill
	\subfigure[{\small Shape-agnostic fabric material identification.}]{%
		\includegraphics[width=0.4\linewidth]{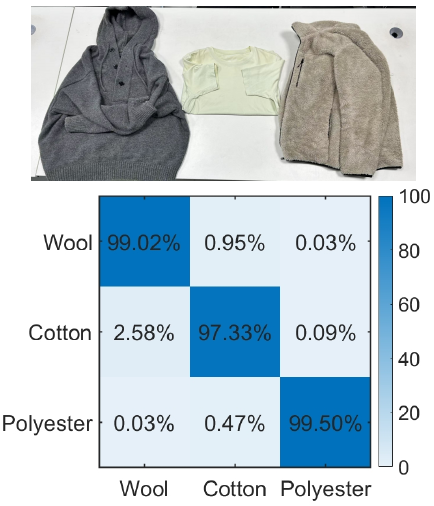}%
		\label{casecloth}%
	}
	\vspace{-5pt}
	\caption{\sysname has potential to be integrated into embodied applications such as cup handling and automate laundry sorting in real-world scenarios. }
	\label{fig:case}
\end{figure}
\subsection{Application Cases}
Accurate material identification has the potential to be integrated into and efficient robotic operations. Traditional robotic perception relies primarily on tactile sensors~\cite{tactilesensing} or electronic skin~\cite{electronicskin}. Robots have to first touch the object to perceive its properties, which may damage fragile items. In contrast, \sysname provides a non-contact identification that is robust to geometry variations, allowing an embodied system to plan its manipulation strategy before physical interaction occurs. We select cup materials and fabrics as representative, application-driven case studies, mirroring common household tasks such as safe glass hand-overs~\cite{handovers} and autonomous laundry sorting~\cite{laundrysorting}. In this section, we extended our pre-trained 20-class model by fine-tuning only the final classification head, evaluating the transferability of our geometry-agnostic features to task-specific categories.

\sssec{Case 1: Orientation-Agnostic Cup Handling.} Robots frequently interact with drinkware where a `one-size-fits-all' grip leads to failure—either crushing a paper cup or dropping a heavy ceramic one. In this application case, we evaluated three common  cup materials: glass, paper, and ceramic. Since a robot may approach a cup from any direction - be it the smooth surface, the rim, or the handle, we tested on entirely unseen orientations. As shown in Fig.~\ref{casecup}, \sysname achieves a high classification accuracy of 98.87\%. This precision enables a robot to distinguish a fragile glass vessel from a deformable paper cup, facilitating context-aware manipulation that prevents both breakage and structural deformation.

\sssec{Case 2: Shape-Agnostic Fabric Sorting.} In autonomous laundry applications, such as sorting or folding clothes, robot must distinguish between fabrics (e.g., sorting wool to prevent shrinkage). However, asingle garment can take infinite forms (i.e., shapes)—neatly folded, spread flat on a table, or crumpled in a basket. In this application, we evaluated materials (e.g.,wool, cotton, and polyester fabrics) that demand distinct laundry and folding strategies in various states of deformation and report the average accuracy. \sysname successfully identifies the material with 98.62\% accuracy (as shown in Fig.~\ref{casecloth}), demonstrating that its signatures are shape-agnostic. Such high-fidelity perception allows embodied systems to automate downstream tasks.
\vspace{-5pt}
\section{Discussion And Limitations}
\sssec{From discrete labels to physical properties:}
While \sysname relies on discrete labels—requiring data collection for new materials—its ability to capture EM and acoustic features enables the potential to estimate continuous physical properties such as complex permittivity and acoustic impedance. By reformulating identification as regression task, \sysname may characterize unseen materials through physical signatures, enabling finer-grained material awareness without retraining, which is left as future work.

\sssec{Multi-object identification:} Currently, \sysname operates under the assumption of a single target object. \sysname aims to tackle more complex scenarios such as multi-object scenarios in the future iterations. Spatial separation techniques, such as beamforming~\cite{beamforming} and blind source separation~\cite{multitarget} has the potential to decompose overlapping signals spatially and enable simultaneous identification.

\sssec{Recognition in fast-motion scenarios:} While \sysname performs robust in static scenarios and support real-time inference with detecting rate of 10$Hz$. However, rapid motions may introduce Doppler-induced distortions. These effects can significantly change the spectral and temporal characteristics, degrading the stability of learned representations. One possible solution is to leverage compensation techniques such as motion-adaptive feature learning~\cite{doppleradaptive} to jointly estimate material and motion parameters, which can potentially extend the framework to fast-moving environments. 
\section{Conclusion}

We present \sysname, a geometry-agnostic multimodal material identification system that combines mmWave and acoustic sensing for robust material identification under varying sensing conditions and surface geometries. By leveraging the geometric consistency shared across the two modalities, \sysname uses cross-modal subtractive disentanglement to suppress geometry-induced factors and extract material-specific representations. We further enhance robustness through contrastive refinement and introduce a pairing-based adaptation strategy for few-shot cross-device generalization. \sysname is evaluated on 20 common materials, showing robustness under diverse distances, orientations and shapes. We also show promising applicability to embodied tasks such as cup handling and fabric sorting.

\bibliographystyle{ACM-Reference-Format}
\bibliography{ref}
\clearpage
\begin{appendices}
\setcounter{table}{0}   
\setcounter{figure}{0}
\renewcommand{\thetable}{A\arabic{table}}
\renewcommand{\thefigure}{A\arabic{figure}}
\section{Lightweight Network}
\label{app:lightweightnetwork}
\begin{figure}[!htbp]
    \centering
    \includegraphics[width=\columnwidth]{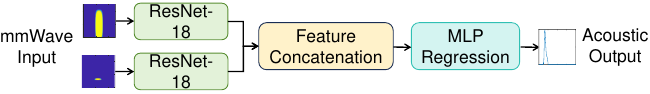}
    \caption{The architecture of lightweight network.}
    \label{fig:lightnet}
\end{figure}
In Section~\ref{sec:DominantShared GeometryVerification}, we adopt a lightweight neural network to validate the dominant geometric information shared across modalities. The network architecture is illustrated in Fig.~\ref{fig:lightnet}. Specifically, ResNet-18~\cite{resnet} is used as the feature extractor for the mmWave modality, with features independently extracted from both the 2D-AoA map and the Range–Azimuth map. The resulting mmWave feature representations are then concatenated to form a complete mmWave embedding, which is fed into a multilayer perceptron to regress the acoustic CIR sequence.

The results are presented in Fig.~\ref{fig:cross_modal_mse}. We observe that performance variations caused by material changes are substantially smaller than those induced by geometric variations, confirming that shared geometric information plays a dominant role in cross-modal prediction.
\section{Savitzky–Golay Filtering}
\label{app:sgfiltering}
To visualize the feature trends in Section~\ref{sssec:scaling}, we employ Savitzky–Golay (SG) filtering~\cite{sgsmoothing} as a post-processing technique. SG filtering performs local polynomial regression within a sliding window, effectively reducing high-frequency noise while preserving the intrinsic waveform morphology and trend of the original features. This enables clearer visualization of feature evolution without altering the underlying semantic structure.

We intentionally use linear layers as the output head instead of convolution-based smoothing modules in the model architecture. Convolution operations introduce local coupling across feature dimensions, which may cause semantic entanglement. As a result, smoothing is applied only for visualization in Fig.~\ref{fig:scale}, and is not incorporated into the model training or inference pipeline.

\section{Loss Function}
\subsection{Barlow Twins Loss}
\label{app:btloss}
To explicitly align features across modalities in \sysname{}, we adopt the Barlow Twins loss~\cite{btloss}. It encourages each embedding dimension to capture semantically meaningful information while enforcing cross-modal correspondence.

Let $\mathbf{z}^{mm}, \mathbf{z}^{ac} \in \mathbb{R}^{B \times D}$ denote the embeddings of $B$ samples from the mmWave and acoustic modalities, where $D$ is the embedding dimension. After normalizing each feature dimension to zero mean and unit variance across the batch, we compute the empirical cross-correlation matrix:

\begin{equation}
\mathbf{C}_{ij} = \frac{1}{B} \sum_{b=1}^{B} z^{mm}_{b,i} \, z^{ac}_{b,j}\nonumber
\end{equation}

The Barlow Twins objective is then formulated as:

\begin{equation}
\mathcal{L}_{BT} =
\sum_{i} (1 - \mathbf{C}_{ii})^2
+ \lambda
\sum_{i} \sum_{j \neq i} \mathbf{C}_{ij}^2\nonumber
\label{eq:barlow}
\end{equation}
where $\lambda$ balances the invariance term (diagonal) and the redundancy reduction term (off-diagonal). The first term enforces \textit{cross-modal invariance}, aligning corresponding semantic dimensions between mmWave and acoustic embeddings. The second term decorrelates different dimensions, preventing redundancy and encouraging each dimension to capture distinct semantic information.

In \sysname{}, this loss ensures that multimodal embeddings share a semantically aligned representation space, mapping corresponding material features from different modalities to the same dimensions.
\subsection{Feature Orthogonality Constraint}
\label{app:corrloss}

To encourage statistical independence between material and geometry embeddings, we define a feature orthogonality loss. Given a material embedding $\mathbf{h}^\mathcal{M}$ and a geometry embedding $\mathbf{h}^\mathcal{G}$, the loss is computed as the cosine similarity between their L2-normalized vectors:

\begin{equation}
\mathcal{L}_{\mathrm{corr}}
=
\frac{{\mathbf{h}^\mathcal{M}}^\top \mathbf{h}^\mathcal{G}}
{\|\mathbf{h}^\mathcal{M}\|_2 \, \|\mathbf{h}^\mathcal{G}\|_2}.
\end{equation}
Minimizing $\mathcal{L}_{\mathrm{corr}}$ penalizes correlation between the embeddings, enforcing orthogonality between the material and geometry subspaces. This encourages subspace disentanglement, ensuring that the material subspace captures only material properties, while the geometry subspace captures only spatial or structural information. In practice, the loss is averaged over batch during training.
\subsection{InfoNCE Loss}
\label{app:infonce}

To mitigate residual cross-modal misalignment after differential processing, \sysname{} employs the InfoNCE loss~\cite{infoNCE} as a contrastive learning objective. This loss encourages embeddings of the same material collected from different locations to cluster together, while pushing embeddings of different materials apart, effectively suppressing spatially varying residual discrepancies and strengthening material representations.

Formally, given a batch of $B$ material embeddings $\{\mathbf{h}^\mathcal{M}_i\}_{i=1}^B$ with material labels $\{y_i\}_{i=1}^B$, the loss is defined as:

\begin{equation}
\mathcal{L}_{con} = -\frac{1}{B} \sum_{i=1}^{B} \log \frac{\sum_{j} \exp\left( \mathbf{h}^\mathcal{M}_i \cdot \mathbf{h}^\mathcal{M}_j / \tau \right) \, \mathbb{I}[y_i = y_j]}{\sum_{k} \exp\left( \mathbf{h}^\mathcal{M}_i \cdot \mathbf{h}^\mathcal{M}_k / \tau \right)}\nonumber
\end{equation}
where $\mathbb{I}[\cdot]$ selects positive pairs (samples of the same material), and $\tau$ is the temperature scaling the logits.  Minimizing $\mathcal{L}_{con}$ suppresses residual noise and misalignment in the material feature space via implicit multi-sample averaging. This ensures that embeddings of the same material remain consistent across locations while maintaining clear separation between different materials, yielding robust material representations under geometric or environmental variations.
\end{appendices}

\end{document}